\documentclass[pdflatex,sn-mathphys-num]{sn-jnl}

\usepackage{xcolor}
\usepackage{soul}
\sethlcolor{yellow}
\soulregister{\cite}{7}
\soulregister{\ref}{7}
\soulregister{\eqref}{7}
\soulregister{\label}{7}
\soulregister{\url}{7}

\usepackage{graphicx}
\usepackage{amsmath,amssymb,amsfonts}
\usepackage{amsthm}
\usepackage{mathtools}
\usepackage{mathrsfs}
\usepackage[title]{appendix}
\usepackage{placeins}
\usepackage{xurl}
\usepackage{enumitem}
\usepackage{booktabs}
\usepackage{tabularx}
\usepackage{xcolor}
\usepackage{textcomp}
\usepackage{manyfoot}
\newcolumntype{Y}{>{\raggedright\arraybackslash}X}
\usepackage{etoolbox}
\hypersetup{hypertexnames=false}

\makeatletter
\patchcmd{\@maketitle}{Corresponding author(s). E-mail(s): }{E-mail: }{}{\ClassError{sn-jnl-patch}{e-mail line patch failed}{}}
\patchcmd{\@maketitle}{\removelastskip\vskip24pt}{\removelastskip\vskip12pt}{}{}
\patchcmd{\@maketitle}{\removelastskip\vskip24pt}{\removelastskip\vskip12pt}{}{}
\newcount\authemailcnt
\def\email#1{\global\advance\emailcnt by 1\relax%
\if@corauemail
   \g@addto@macro\corrauthemail{\setcounter{footnote}{0}\textcolor{blue}{#1}}%
\else
   \ifnum\authemailcnt>0 \g@addto@macro\authemail{; }\fi
   \global\advance\authemailcnt by 1\relax
   \g@addto@macro\authemail{\setcounter{footnote}{0}\textcolor{blue}{#1}}%
\fi}
\makeatother

\newcommand{\vE}{\mathbf{E}}
\newcommand{\vB}{\mathbf{B}}

\newcommand{\vJ}{\mathbf{J}}
\newcommand{\vK}{\mathbf{K}}
\newcommand{\vA}{\mathbf{A}}
\newcommand{\vS}{\mathbf{S}}

\newcommand{\vr}{\mathbf{r}}

\newcommand{\nhat}{\hat{\mathbf{n}}}

\newcommand{\zhat}{\hat{\mathbf{z}}}
\newcommand{\eps}{\varepsilon_{0}}
\newcommand{\muz}{\mu_{0}}
\newcommand{\curl}{\nabla\times}
\newcommand{\dive}{\nabla\cdot}
\newcommand{\grad}{\nabla}
\newcommand{\pdt}[1]{\frac{\partial #1}{\partial t}}

\newcommand{\Et}{\mathbf{E}_{t}}
\newcommand{\Bt}{\mathbf{B}_{t}}
\newcommand{\Ez}{E_{z}}
\newcommand{\Bz}{B_{z}}
\newcommand{\gradt}{\nabla_{\!t}}
\newcommand{\lapt}{\nabla_t^{2}}
\newcommand{\kc}{\gamma}
\newcommand{\cA}{\mathcal A}
\newcommand{\bnd}{S}

\newcommand{\avg}[1]{\left\langle #1\right\rangle}
\theoremstyle{thmstyleone}
\newtheorem{theorem}{Theorem}

\newtheorem{proposition}{Proposition}
\newtheorem{corollary}{Corollary}
\theoremstyle{thmstyletwo}

\allowdisplaybreaks

\begin{document}

\title[Scalar-Longitudinal Modes of Extended Electrodynamics in Perfectly Conducting Waveguides and Divergence Cleaning Interpretation]
{Scalar-Longitudinal Modes of Extended Electrodynamics in Perfectly Conducting Waveguides and Divergence Cleaning Interpretation}

\author*[1]{\fnm{N.} \sur{Rentzber}}\email{nrentzbe@uccs.edu}
\author[1]{\fnm{J.} \sur{Howard}}
\author[2,3]{\fnm{F. O.} \sur{Minotti}}\email{minotti@df.uba.ar}
\author[1]{\fnm{A.} \sur{Pinchuk}}
\author[4]{\fnm{G.} \sur{Modanese}}\email{giovanni.modanese@unibz.it}

\affil*[1]{\orgdiv{Department of Physics and Energy Science}, 

\orgname{University of Colorado}, \orgaddress{\city{Colorado Springs}, \postcode{80918}, \country{USA}}}

\affil[2]{\orgname{Universidad de Buenos Aires, Facultad de Ciencias Exactas y Naturales, Departamento de F\'{\i}sica}, \orgaddress{\city{Buenos Aires}, \country{Argentina}}}

\affil[3]{\orgname{CONICET-Universidad de Buenos Aires, Instituto de F\'{\i}sica Interdisciplinaria y Aplicada (INFINA)}, \orgaddress{\city{Buenos Aires}, \country{Argentina}}}

\affil[4]{\orgdiv{Faculty of Engineering}, 

\orgname{Free University of Bozen-Bolzano}, \orgaddress{\city{Bolzano}, \postcode{39100}, \country{Italy}}}

\abstract{\unboldmath Extended electrodynamics keeps the scalar combination $C=\dive\vA+\ensuremath{\muz}\eps\,\partial\Phi/\partial t$ as a dynamical field rather than imposing the Lorenz gauge condition, and this scalar adds a scalar-longitudinal (SL) family to the TE, TM and TEM modes. With $C=0$ on the wall of a closed perfectly conducting waveguide, the classical modes are unchanged. The SL modes have $\mathbf B=0$ and an electric field along $\grad C$, and their cutoffs are the Dirichlet eigenvalues of the cross-section, the same as the TM cutoffs, with no coaxial TEM counterpart. Charge-conserving sources with zero scalar data do not excite them. Matching at a conductor described by its constitutive laws gives $\partial C/\partial n=0$ instead and couples the scalar-bearing modes to the Maxwell field. The results belong to the $C=0$ problem, the condition that separates the sectors and the one divergence cleaning uses, with benchmarks for the rectangular, circular cylinder, coaxial and L-shaped guides.}


\maketitle

\clearpage
\setcounter{tocdepth}{2}
{\small\tableofcontents}
\clearpage

\section{Introduction}
\label{sec:intro}
Classical waveguide theory decomposes the source-free field in a hollow perfectly conducting guide into transverse-electric (TE), transverse-magnetic (TM) and, on multiply connected cross-sections, transverse-electromagnetic (TEM) traveling modes. This decomposition is foundational for microwave engineering and guided-wave theory~\cite{jackson,pozar,collin}.
\\

No claim is made that the scalar-longitudinal sector is realized in nature. Local charge conservation, whose failure is the only source of $C$, is among the best-tested principles in physics~\cite{okun,borexino,majorana}. The aim is to say exactly what the extended framework predicts in a theoretical closed conducting guide.

\subsection{Classical Electrodynamics}
In classical electrodynamics, meaning standard Maxwell theory throughout this paper, the macroscopic field equations in vacuum with total charge density $\rho_{\mathrm{tot}}$ and total current density 
$\vJ_{\mathrm{tot}}$ are

\begin{equation}
\begin{aligned}
\dive\vE &= \frac{\rho_{\mathrm{tot}}}{\eps},
\qquad&
\curl\vE &= -\pdt{\vB},
\\[1mm]
\dive\vB &= 0,
\qquad&
\curl\vB &= \mu_{0}\vJ_{\mathrm{tot}}
+\mu_{0}\epsilon_{0}\pdt{\vE}.
\end{aligned}
\label{eq:maxwell}
\end{equation}
\medskip

The divergence of Amp\`ere's law and the time derivative of Gauss's law gives the continuity equation,

\begin{equation}
\pdt{\rho_{\mathrm{tot}}}+\dive\vJ_{\mathrm{tot}}=0.
\label{eq:continuity}
\end{equation}
\medskip

Classical electrodynamics forces local conservation of the total charge. A source pair with a nonzero anomaly

\begin{equation}
\Lambda\equiv\pdt{\rho_{\mathrm{tot}}}+\dive\vJ_{\mathrm{tot}}
\label{eq:anomaly}
\end{equation}
\\
has no Maxwell solution.

\subsection{Potentials and the Lorenz Gauge}
The homogeneous pair in Eq.~\eqref{eq:maxwell} is solved identically by potentials,

\begin{equation}
\vB=\curl\vA,
\qquad
\vE=-\grad\Phi-\pdt{\vA},
\label{eq:potentials}
\end{equation}
\medskip

Substituting Eq.~\eqref{eq:potentials} into the inhomogeneous pair and using $\curl\curl\vA=\grad(\dive\vA)-\nabla^{2}\vA$ gives the coupled potential equations
\begin{equation}
\nabla^{2}\vA-\mu_{0}\eps\frac{\partial^{2}\vA}{\partial t^{2}}=-\mu_{0}\vJ_{\mathrm{tot}}+\grad C,
\qquad
\nabla^{2}\Phi-\mu_{0}\eps\frac{\partial^{2}\Phi}{\partial t^{2}}=-\frac{\rho_{\mathrm{tot}}}{\eps}-\pdt{C},
\label{eq:potcoupled}
\end{equation}
in which the potentials are coupled through the scalar
\begin{equation}
C\equiv\dive\vA+\mu_{0}\eps\pdt{\Phi}.
\label{eq:Cdef}
\end{equation}
Choosing the Lorenz gauge, $C=0$, decouples Eq.~\eqref{eq:potcoupled} into the two wave equations
\begin{equation}
\nabla^{2}\vA-\mu_{0}\eps\frac{\partial^{2}\vA}{\partial t^{2}}=-\mu_{0}\vJ_{\mathrm{tot}},
\qquad
\nabla^{2}\Phi-\mu_{0}\eps\frac{\partial^{2}\Phi}{\partial t^{2}}=-\frac{\rho_{\mathrm{tot}}}{\eps},
\label{eq:potwave}
\end{equation}

with $c=1/\sqrt{\mu_{0}\eps}$, and the same wave operator governs the fields themselves. Taking the curl of Faraday's law, using Amp\`ere's and Gauss's laws, and the curl of Amp\`ere's law with Faraday's law and $\dive\vB=0$,

\begin{equation}
\nabla^{2}\vE-\mu_{0}\eps\frac{\partial^{2}\vE}{\partial t^{2}}=\frac{\grad\rho_{\mathrm{tot}}}{\eps}+\mu_{0}\pdt{\vJ_{\mathrm{tot}}},
\qquad
\nabla^{2}\vB-\mu_{0}\eps\frac{\partial^{2}\vB}{\partial t^{2}}=-\mu_{0}\curl\vJ_{\mathrm{tot}},
\label{eq:fieldwave}
\end{equation}
which in a source-free region are the vacuum wave equations. A gauge transformation can make $C$ nonzero without changing $\vE$ or $\vB$.

\subsection{Extended Electrodynamics}
Extended electrodynamics is defined by one postulate. The potentials obey the wave equations \eqref{eq:potwave} exactly as in the Lorenz gauge, but the combination $C$ of Eq.~\eqref{eq:Cdef} is not set to zero and is kept as a field. It is this extended Lorenz condition, and not the wave equations, that changes and extends Maxwell's equations. Using Eq.~\eqref{eq:potentials} together with the wave equations~\eqref{eq:potwave} and retaining $C$ gives the extended Maxwell's equations

\begin{equation}
\begin{gathered}
\dive\vE
=\frac{\rho_{\mathrm{tot}}}{\eps}
-\frac{\partial C}{\partial t},
\qquad
\curl\vE
=-\frac{\partial\vB}{\partial t},
\\[1mm]
\dive\vB=0,
\qquad
\curl\vB
=\mu_{0}\vJ_{\mathrm{tot}}
+\mu_{0}\eps\frac{\partial\vE}{\partial t}+\grad C.
\end{gathered}
\label{eq:eedfields}
\end{equation}
\medskip

Amp\`ere's law acquires $\grad C$, while Gauss's law acquires $\partial C/\partial t$, and both new terms come from extending the Lorenz gauge, which therefore extends Maxwell's equations. Setting $C\equiv0$ in the same substitution gives back Eq.~\eqref{eq:maxwell}, which is how the classical electrodynamics theory sits inside the extended one. Where Maxwell theory forced Eq.~\eqref{eq:continuity}, the divergence of the extended Amp\`ere's law added to the time derivative of the extended Gauss's law that now gives a wave equation,
\begin{equation}
\nabla^{2}C-\mu_{0}\eps\frac{\partial^{2}C}{\partial t^{2}}=-\mu_{0}\Lambda,
\label{eq:Cwave}
\end{equation}
so the scalar is sourced only by the anomaly of Eq.~\eqref{eq:anomaly}. The field-form wave equations are unchanged. Repeating the curl computations that gave Eq.~\eqref{eq:fieldwave} with the extended system, the terms in $C$ cancel identically and Eq.~\eqref{eq:fieldwave} holds as written, which is why the scalar can hide inside Maxwell theory. It enters the fields only through equations \eqref{eq:eedfields}, and in a source-free region all three of $\vE$, $\vB$ and $C$ obey the same vacuum wave equation.
\\

In classical electrodynamics the Lorenz condition sets $C=0$ and the scalar carries no physics. In extended electrodynamics, it behaves as a field. The formulation goes back to Ohmura~\cite{ohmura} and to Aharonov and Bohm~\cite{aharonovbohm}. The Stueckelberg construction~\cite{stueckelberg,stueckelberg2} contains it in a formally related form, and it has been developed by Van Vlaenderen and Waser~\cite{vanvlaenderen}, Hively and Giakos~\cite{hively2012}, Woodside~\cite{woodside1999,woodside2009}, and others~\cite{hivelyloebl,modanese2017,minotti2021,reedhively2020,jimenezmaroto}. The formulation followed here, with its energy law in Section~\ref{sec:extenergy} and the wall condition at a physical conductor in Section~\ref{sec:extwall}, is that of Minotti and Modanese~\cite{minotti2021,minotti2023}. In a closed resonator the independently imposed condition $C=0$ gives the Dirichlet spectrum of the volume, and every guide closed by two end walls carries the TM modes with axial index $p\geq1$ as scalar modes. This paper solves the long uniform guide, asking throughout what changes when the scalar is kept, mode by mode and cutoff by cutoff.
\\

This same scalar extension also arises for a different reason in computational electromagnetics. A Maxwell solver preserves Gauss's law when its grid operators satisfy the divergence-of-curl identity, its sources conserve charge on the grid and its initial fields satisfy the law, as in schemes that compute the discrete current in a form compatible with the discrete Maxwell solver~\cite{crouseilles2014}. The same operator identity allows Faraday's law to preserve $\dive\vB=0$ when that constraint holds initially. In schemes that do not preserve Gauss's law, errors in $\dive\vE$ can accumulate and corrupt the solution. Hyperbolic divergence cleaning controls these errors by introducing an auxiliary scalar, coupled through a gradient term in Amp\`ere's law and a time derivative in Gauss's law in the same way as $C$ in Eq.~\eqref{eq:eedfields}. The scalar carries divergence error as a wave at a chosen speed, with damping used to reduce it~\cite{munz2000,dedner2002}. Without damping and with the cleaning speed set to $c$, the auxiliary equations coincide with the extended system. In a closed conducting guide with the same scalar wall condition, the cleaning scalar therefore has the modes derived here. These modes provide benchmarks for checking a solver's field profiles, cutoffs and convergence. Section~\ref{sec:geom} gives closed-form solutions for the three standard guides, while the L-shaped guide of Section~\ref{sec:lshape} provides a numerical benchmark for a non-separable geometry.

\subsection{Closed Waveguide}
Sections~\ref{sec:classical} and~\ref{sec:eed} set the two theories in the closed waveguide, Section~\ref{sec:slsector} isolates the scalar-longitudinal sector and shows that the wall closes it, Section~\ref{sec:theorem} proves the spectral theorem, and Section~\ref{sec:geom} solves the rectangular, circular, coaxial and L-shaped guides for every family. The remaining sections ask what any new mode family has to answer, where it cuts off, what it carries, how it behaves at a step and how it would be driven, before Section~\ref{sec:cleaning} states the correspondence with divergence cleaning in full.
\\

With $C=0$ on the wall the SL modes of a guide with cross-section $\cA$ are the eigenmodes of the Dirichlet Laplacian on $\cA$, with $\vB=\mathbf0$, $\vE$ along the gradient of $C$, and cutoff frequencies equal to those of the TM modes,
\begin{equation}
\boxed{-\lapt\phi_j=\lambda_j\phi_j,
\quad
\phi_j|_{\bnd}=0,
\qquad
\omega_{c,j}^{\mathrm{SL}}=c\sqrt{\lambda_j}=\omega_{c,j}^{\mathrm{TM}}},
\label{eq:introresult}
\end{equation}
\medskip

\noindent For every propagating mode of the scalar sector alone, the conducting wall requires $C=0$. With this condition, every SL mode has a nonzero cutoff, and no propagating scalar branch exists below the first TM cutoff. In the canonical guides, the SL and TM cutoffs coincide mode for mode, so the scalar introduces no new cutoff frequencies and provides no counterpart to the coaxial TEM mode. Each propagating SL mode has equal electric and scalar contributions to the conserved quadratic form of Section~\ref{sec:extenergy} and ends on wall charge without a parallel wall current. Within the good-conductor loss model used here, its vanishing magnetic field eliminates the associated wall-loss mechanism, while its TM partner is attenuated. At a step, the matching conditions involve only $C$ and its derivative along the axis. Conserved sources leave the scalar sector empty when the scalar initial data vanish and no scalar field enters through the boundaries. Finally, undamped divergence cleaning at speed $c$ obeys the same equations with the same scalar wall condition, making each SL mode an exact benchmark for these numerical solvers. At a conductor described by its constitutive laws, the matching of Ref.~\cite{minotti2023} gives $\partial C/\partial n=0$ on the vacuum side instead. Under that condition no purely SL mode propagates in a uniform guide, and the scalar-bearing modes are coupled to the Maxwell field, a problem left for a separate study.

\section{Classical Electrodynamics in the Closed Guide}
\label{sec:classical}
 
Everything here is Maxwell theory, in the form Section~\ref{sec:eed} deforms, so that every later comparison is between displayed equations. The guide is filled with vacuum, with permittivity $\eps$, permeability $\mu_{0}$ and wave speed $c=1/\sqrt{\mu_{0}\eps}$, and a lossless dielectric fill follows by replacing $\eps$ and $\mu_{0}$ with $\varepsilon$ and $\mu$. Let $\cA\subset\mathbb R^2$ be the bounded, connected cross-section of a guide uniform along $z$, with piecewise smooth boundary and perfectly conducting side wall $\bnd$. The cross-section may be multiply connected, as in a coaxial line.
 
\subsection{Parallel and Perpendicular Wall Conditions}
On the perfectly conducting wall $S$ the parallel electric field and the perpendicular magnetic field vanish,

\begin{equation}
\nhat\times\vE=\mathbf0,
\qquad
\nhat\cdot\vB=0
\qquad\text{on } S.
\label{eq:wallclassical}
\end{equation}

Resolved along the guide, the first condition is $\Ez=0$ together with $\hat{\mathbf t}\cdot\Et=0$, where $\hat{\mathbf t}$ is the unit tangent of $S$ in the cross-section. For fields of this form the reduction relations of the next subsection turn these into the classical boundaries of Eq.~\eqref{eq:classicaldata}. Dirichlet data for $\Ez$ follow directly from the axial component of $\nhat\times\vE=\mathbf0$. For TE fields, $\Et=-(i\omega/\gamma^{2})\,\zhat\times\gradt\Bz$, and with $\hat{\mathbf t}=\zhat\times\nhat$ the parallel gives $\hat{\mathbf t}\cdot\Et=-(i\omega/\gamma^{2})\,\partial\Bz/\partial n=0$, so $\partial\Bz/\partial n=0$ on $S$, including at cutoff. The condition $\nhat\cdot\vB=0$ is then a consequence rather than the source of the Neumann data, and for TM fields $\nhat\cdot\Bt$ is proportional to the derivative of $\Ez$ along $S$, which vanishes because $\Ez$ does.
 
\subsection{Energy}
Classically, the Poynting theorem is defined as
\begin{equation}
\frac{\partial u_{M}}{\partial t}+\dive\vS_{M}=-\vJ_{\mathrm{tot}}\cdot\vE,
\qquad
u_{M}=\frac{\eps}{2}|\vE|^{2}+\frac{1}{2\mu_{0}}|\vB|^{2},
\qquad
\vS_{M}=\frac{1}{\mu_{0}}\vE\times\vB.
\label{eq:poyntingM}
\end{equation}

In a source-free guide with perfectly conducting walls, $\nhat\times\vE=0$ removes the wall flux, the total energy is conserved, and the positive definite density gives uniqueness of the time-dependent problem. The extended counterpart is Proposition~\ref{prop:uniqueness}, and setting $C=0$ there recovers the classical statement.
 
\subsection{Classical Guide Problem}
Every field is written as a complex exponential traveling along $z$, for example $\vE(x,y,z,t)=\vE(x,y)\,e^{i(kz-\omega t)}$, with the physical field understood as the real part. Differentiation then acts as multiplication,

\begin{equation}
\grad=\gradt+ik\zhat,
\qquad
\frac{\partial}{\partial t}=-i\omega,
\qquad
\frac{\partial}{\partial z}=ik.
\end{equation}
The transverse wavenumber is
\begin{equation}
\kc^2=\omega^{2}\mu_{0}\eps-k^2,
\qquad
\omega_c=c\kc.
\label{eq:disp}
\end{equation}

The axial wavenumber $k$ and the cutoff eigenvalue $\gamma^{2}$ follow the notation of Ref.~\cite{jackson}. With this dependence and the transverse wavenumber of Eq.~\eqref{eq:disp}, the axial components satisfy
\begin{equation}
\left(\lapt+\kc^{2}\right)\Ez=0,
\qquad
\left(\lapt+\kc^{2}\right)\Bz=0,
\label{eq:axialhelm}
\end{equation}
and the perfect-conductor condition $\nhat\times\vE=0$ then selects
\begin{equation}
\Ez|_{\bnd}=0\ \ \text{(TM)},
\qquad
\partial\Bz/\partial n|_{\bnd}=0\ \ \text{(TE)},
\qquad
\kc=0\ \ \text{(TEM)}.
\label{eq:classicaldata}
\end{equation}
Equations~\eqref{eq:maxwell} to~\eqref{eq:classicaldata} give the familiar TE, TM and, on multiply connected cross-sections, TEM modes. For $\kc\ne0$, the transverse fields follow from the axial fields. Eliminating $\Et$ and $\Bt$ between the two curl equations of the source-free guide gives the classical relations
\begin{align}
\Et&=\frac{i}{\kc^{2}}\left[k\,\gradt\Ez-\omega\,\zhat\times\gradt\Bz\right],
\label{eq:transM}\\
\Bt&=\frac{i}{\kc^{2}}\left[k\,\gradt\Bz+\omega\mu_{0}\eps\,\zhat\times\gradt\Ez\right],
\label{eq:transBM}
\end{align}
The axial field generators satisfy the transverse Helmholtz equation of Eq.~\eqref{eq:axialhelm}, and the perfect-conductor condition selects the classical data of Eq.~\eqref{eq:classicaldata}. TM modes carry Dirichlet data for $\Ez$, TE modes carry Neumann data for $\Bz$, and the singular case $\kc=0$ is the TEM family, which requires at least two conducting boundary components and is built from a harmonic transverse potential with one voltage potential per component~\cite{jackson,pozar,collin}. The classical cutoffs are therefore the Dirichlet spectrum of the cross-section for TM modes, the nonzero Neumann spectrum for TE modes, and $\kc=0$ for the $q-1$ TEM modes of a cross-section with $q$ conductor boundary components. A general classical field is a sum over these modal families, and the summation expansions of Section~\ref{sec:geom} and the junction analysis of Section~\ref{sec:matching} use this structure.
 
\section{Extended Electrodynamics in the Closed Guide}
\label{sec:eed}
 
The extended system of Eq.~\eqref{eq:eedfields} is placed in the same guide, with $\rho_{\mathrm{tot}}=0$ and $\vJ_{\mathrm{tot}}=\mathbf0$ in the interior, so that $C$ obeys the homogeneous wave equation and enters only through the constraint equations and the wall.
 
\subsection{Extended Energy Law}
\label{sec:extenergy}
The field equations imply the following quadratic balance identity,
\begin{equation}
\frac{\partial u}{\partial t}+\dive\vS=-\vJ_{\mathrm{tot}}\cdot\vE+\frac{\rho_{\mathrm{tot}}C}{\mu_{0}\eps},
\label{eq:energy}
\end{equation}
where

\begin{equation}
u=\frac{\eps}{2}|\vE|^{2}+\frac{1}{2\mu_{0}}|\vB|^{2}+\frac{1}{2\mu_{0}}C^{2},
\qquad
\vS=\frac{1}{\mu_{0}}\vE\times\vB+\frac{1}{\mu_{0}}C\vE.
\label{eq:density}
\end{equation}
\medskip

\noindent For a source-free guide, the right side of Eq.~\eqref{eq:energy} vanishes. The scalar adds a positive term to $u$ and a term along $\vE$ to $\vS$. Therefore $\vS$ rather than the Maxwell Poynting vector of Eq.~\eqref{eq:poyntingM} alone is conserved in the extended system, and the term $C\vE/\mu_{0}$ is the part of $\vS$ that the magnetic-field-free SL modes of Section~\ref{sec:theorem} carry.
\\

Equation~\eqref{eq:energy} does not fix $u$ and $\vS$. For any vector fields $\mathbf{X}$ and $\mathbf{Y}$,
\begin{equation}
u'=u+\dive\mathbf{X},\qquad \vS'=\vS-\pdt{\mathbf{X}}+\curl\mathbf{Y}
\label{eq:XY}
\end{equation}
satisfies the same balance. The energy law follows from the Lagrangian in Section~5 of Ref.~\cite{minotti2021}, with the flux in its Eq.~(39) and the balance in its Eq.~(40). The density and flux are also quoted in Eqs.~(2) and~(3) of Ref.~\cite{minotti2023}. The law reads
\begin{equation}
\begin{gathered}
\pdt{u_{L}}+\dive\vS_{L}=-\vJ_{\mathrm{tot}}\cdot\vE+\Lambda\,\Phi,\\[2pt]
u_{L}=\frac{\eps}{2}|\vE|^{2}+\frac{1}{2\mu_{0}}|\vB|^{2}+\eps\,\Phi\,\pdt{C}-\frac{1}{\mu_{0}}\,\vA\cdot\grad C-\frac{1}{2\mu_{0}}\,C^{2},\\[2pt]
\vS_{L}=\frac{1}{\mu_{0}}\Big(\vE\times\vB-\Phi\,\grad C+\vA\,\pdt{C}\Big),
\end{gathered}
\label{eq:Lpair}
\end{equation}
with $\Lambda$ the anomaly of Eq.~\eqref{eq:Cwave}. Its source term $\Lambda\Phi$ replaces $\rho_{\mathrm{tot}}C/(\mu_{0}\eps)$ of Eq.~\eqref{eq:energy}, which Eq.~\eqref{eq:XY} cannot change. The potential wave equations give, as in Section~6 of Ref.~\cite{minotti2021},
\begin{equation}
\Lambda\Phi-\frac{\rho_{\mathrm{tot}}C}{\mu_{0}\eps}=\pdt{}\Big[\eps\Big(\Phi\pdt{C}-C\pdt{\Phi}\Big)\Big]-\dive\Big[\frac{1}{\mu_{0}}\big(\Phi\grad C-C\grad\Phi\big)\Big],
\label{eq:sourcediff}
\end{equation}
and with $\dive\vA=C-\mu_{0}\eps\,\partial\Phi/\partial t$,
\begin{equation}
\begin{gathered}
u_{L}=u+\eps\Big(\Phi\pdt{C}-C\pdt{\Phi}\Big)-\frac{1}{\mu_{0}}\dive(C\vA),\\[2pt]
\vS_{L}=\vS-\frac{1}{\mu_{0}}\big(\Phi\grad C-C\grad\Phi\big)+\frac{1}{\mu_{0}}\pdt{}(C\vA).
\end{gathered}
\label{eq:LfromF}
\end{equation}
Every term that separates the two pairs contains $\Phi$ or $\vA$.
\\

A residual gauge transformation,
\begin{equation}
\vA\to\vA+\grad\zeta,\qquad \Phi\to\Phi-\pdt{\zeta},\qquad \nabla^{2}\zeta-\mu_{0}\eps\,\frac{\partial^{2}\zeta}{\partial t^{2}}=0,
\label{eq:gauge}
\end{equation}
leaves $\vE$, $\vB$ and $C$ unchanged and changes the pair by
\begin{equation}
\delta u_{L}=-\eps\,\pdt{\zeta}\,\pdt{C}-\frac{1}{\mu_{0}}\grad\zeta\cdot\grad C,\qquad \delta\vS_{L}=\frac{1}{\mu_{0}}\Big(\pdt{\zeta}\,\grad C+\pdt{C}\,\grad\zeta\Big).
\label{eq:gaugechange}
\end{equation}
Complete source information with retarded solutions removes this freedom, Section~8.1 of Ref.~\cite{minotti2021}. The guide problem fixes $\vE$, $\vB$ and $C$ only, so its field data do not fix $u_{L}$ or $\vS_{L}$. For a propagating SL mode, with $U_{L}$ and $P_{L}$ the cross-section integrals of $\langle u_{L}\rangle$ and $\langle\vS_{L}\cdot\zhat\rangle$,
\begin{equation}
\begin{aligned}
(U_{L},P_{L})&=(U_{j},P_{j}) & &\text{for } \Phi=0,\ \ \vA=-\frac{\grad C}{\omega^{2}\mu_{0}\eps},\\
(U_{L},P_{L})&=(-U_{j},-P_{j}) & &\text{for } \vA=\mathbf0,\ \ \Phi=\frac{iC}{\omega\mu_{0}\eps},
\end{aligned}
\label{eq:gaugemodes}
\end{equation}
with $U_{j}$ and $P_{j}$ from Proposition~\ref{prop:powerenergy}. For a field with $C\neq0$ in a closed guide the sources include the charge and current at the wall, so its physical energy and flux also depend on how the wall supplies its charge, the question left open in Sections~\ref{sec:extwall} and~\ref{sec:wallcharge}. For the Maxwell modes $C=0$, and Eqs.~\eqref{eq:Lpair} and~\eqref{eq:gaugechange} reduce to
\begin{equation}
u_{L}=\frac{\eps}{2}|\vE|^{2}+\frac{1}{2\mu_{0}}|\vB|^{2},\qquad \vS_{L}=\frac{1}{\mu_{0}}\,\vE\times\vB,\qquad \delta u_{L}=0,\qquad \delta\vS_{L}=\mathbf0,
\label{eq:Lmaxwell}
\end{equation}
so their energy and power are the classical ones and are fixed by the fields.
\\

Reference~\cite{minotti2021} derives Eq.~\eqref{eq:energy} as its Eq.~(31) and treats it as a correct relation among the fields rather than an energy law. Here it serves as a conserved quadratic form, positive definite, gauge invariant and equal to the Maxwell pair at $C=0$. Propositions~\ref{prop:uniqueness} and~\ref{prop:powerenergy} to~\ref{prop:impedance} and the evanescent flux of Section~\ref{sec:cutoffs} use it, and energy, flux and power in the scalar sector below mean $u$ and $\vS$. With the complete transverse basis of Corollary~\ref{cor:complete}, the uniqueness below makes the explicit modes a basis for the modal analysis.
\\

\begin{proposition}[Conserved Form and Uniqueness]
\label{prop:uniqueness}
With perfectly conducting walls and $C=0$ on them, no flux of $\vS$ passes through the wall, and a source-free field is fixed by its initial values and its values on the two end cross-sections.
\end{proposition}

\begin{proof}
With no sources Eq.~\eqref{eq:energy} is $\partial u/\partial t+\dive\vS=0$, with $u=\tfrac12\eps|\vE|^{2}+|\vB|^{2}/(2\mu_{0})+C^{2}/(2\mu_{0})$ and $\vS=(\vE\times\vB+C\vE)/\mu_{0}$. On the wall $\vE$ is perpendicular to the surface, so $(\vE\times\vB)\cdot\nhat=0$, and $C=0$, so $\vS\cdot\nhat=0$. The integral of $u$ over a length of guide therefore changes only through its end faces. The difference of two solutions with the same initial and end-face fields starts with that integral at zero and gains none, and $u$ is a sum of squares.
\end{proof}

The result applies to finite-energy fields such as guided wave packets. It also applies to a truncated guide when the end faces carry no net energy flux. Otherwise, the end-flux term must be kept. Monochromatic modes on an infinite guide have infinite total energy and are instead normalized per unit length as in Proposition~\ref{prop:powerenergy}.
 
\subsection{Extended Wall Conditions}
\label{sec:extwall}
The classical wall pair $\nhat\times\vE=\mathbf0$ and $\nhat\cdot\vB=0$ of Section~\ref{sec:classical} is not enough for the extended system, and a count of waves shows why. Along the direction perpendicular to the wall, Eq.~\eqref{eq:eedfields} carries seven field components. Six of them combine into waves traveling at speed $c$, three toward the wall and three away from it, and the seventh, $B_n$, does not travel perpendicular to the wall at all, which is why $\dive\vB=0$ propagates as a constraint rather than needing its own wall condition. The three outgoing waves are fixed by the field inside the guide, so the three incoming ones must be given at the wall~\cite{benzonigavage}. In Maxwell theory there are only two incoming waves and the two components of $\nhat\times\vE=\mathbf0$ supply them. The extended system has a third, the combination $E_n-cC$, and the conductor says nothing about it. One more wall condition on $C$ is needed.
\medskip

That condition is

\begin{equation}
C=0
\qquad\text{on }S,
\label{eq:Cwallchoice}
\end{equation}
\medskip

\noindent Three facts support the choice in Eq.~\eqref{eq:Cwallchoice}. First, Proposition~\ref{prop:bc} shows that the conducting wall requires $C=0$ for every propagating mode of the scalar sector alone in the uniform guide. Second, this condition removes the wall flux of the quadratic form used in the uniqueness proof. Third, the same condition is used for the cleaning scalar at a perfect conductor in Refs.~\cite{munz2000,munzomnesschneider,crestetto2012}. With this boundary choice, the application in Section~\ref{sec:cleaning} requires no physical model of the conductor. Since $C=0$ also gives $\partial C/\partial t=0$ on the wall, the quadratic and Lagrangian fluxes there are
\begin{equation}
\begin{gathered}
\nhat\cdot\vS=\frac{CE_{n}}{\mu_{0}}=0,\\[2pt]
\nhat\cdot\vS_{L}=-\frac{\Phi}{\mu_{0}}\frac{\partial C}{\partial n}
=\begin{cases}
-\Phi\,\Lambda_{s}, & \text{no current in the conductor},\\[2pt]
-\Phi\,\nhat_{c}\cdot\vJ_{f,c}, & \text{perpendicular current in the conductor}.
\end{cases}
\end{gathered}
\label{eq:Lwallflux}
\end{equation}
The two cases for $\vS_{L}$ correspond to the conductor assumptions discussed in Section~\ref{sec:wallcharge}. Proposition~\ref{prop:uniqueness} uses the vanishing flux of $\vS$. The Lagrangian flux depends on the wall potential $\Phi$, which the field data do not determine. The boundary condition therefore closes the quadratic balance without fixing the physical energy exchange with the wall.
\\

In the divergence-cleaning setting, Fornet and Mouysset~\cite{fornetmouysset} show that recovering Maxwell solutions requires compatible scalar initial and boundary data. Their rectangular-guide example has a zero scalar wall value but admits non-Maxwell waves when the scalar initial data are nonzero.
\\

At a physical interface between the vacuum and a conductor, the condition on $C$ is not chosen but obtained by matching. The field equations are applied on each side of the interface, where the constitutive laws are defined, and the two solutions are joined there. The interface conditions are continuity of $\Phi$ and $A_{n}$, continuity of $\vE_{\parallel}$ and $B_{n}$, and the surface charge balance, and no further condition is imposed at the interface. Minotti and Modanese carry out this matching~\cite{minotti2023}, and at perpendicular incidence it is short enough to give in full. The derivation below is that of one of the authors (F.M.), written in the notation of this paper. The conductor is nonmagnetic and nonpolarizable, with conductivity $\sigma_{c}$, locally conserved charge and no anomaly. Let $n$ measure distance along $\nhat$ from the vacuum into the conductor, with the wall at $n=0$, and let every quantity depend on $n$ and $t$ alone, so that $\vA$ has the single component $A_{n}$. In the potentials, with $C$ defined by Eq.~\eqref{eq:Cdef}, the extended system is equivalent to the wave equations of Ref.~\cite{minotti2021}, their Eq.~(30),
\begin{equation}
\frac{\partial^{2}\Phi}{\partial n^{2}}-\mu_{0}\eps\frac{\partial^{2}\Phi}{\partial t^{2}}=-\frac{\rho}{\eps},
\qquad
\frac{\partial^{2}A_{n}}{\partial n^{2}}-\mu_{0}\eps\frac{\partial^{2}A_{n}}{\partial t^{2}}=-\mu_{0}J_{n},
\label{eq:potwaves}
\end{equation}
with $\rho=0$ and $J_{n}=0$ in the vacuum, and in the conductor
\begin{equation}
J_{n}=\sigma_{c}E_{n}=-\sigma_{c}\Big(\frac{\partial\Phi}{\partial n}+\frac{\partial A_{n}}{\partial t}\Big),
\qquad
\frac{\partial\rho}{\partial t}+\frac{\partial J_{n}}{\partial n}=0.
\label{eq:conductorlaws}
\end{equation}
In the vacuum the solution is an incident and a reflected wave, and in the conductor a transmitted one,
\begin{equation}
\begin{aligned}
\Phi&=\Phi_{i}\,e^{i(kn-\omega t)}+\Phi_{r}\,e^{i(-kn-\omega t)}, & A_{n}&=A_{i}\,e^{i(kn-\omega t)}+A_{r}\,e^{i(-kn-\omega t)}, & &n<0,\\
\Phi&=\Phi_{t}\,e^{i(k_{c}n-\omega t)}, & A_{n}&=A_{t}\,e^{i(k_{c}n-\omega t)}, & &n>0,
\end{aligned}
\label{eq:matchwaves}
\end{equation}
with $k=\omega/c$ in the vacuum. Substituting the transmitted wave into Eq.~\eqref{eq:conductorlaws}, with $\partial/\partial n\to ik_{c}$ and $\partial/\partial t\to-i\omega$, gives $J_{n}=i\sigma_{c}(\omega A_{t}-k_{c}\Phi_{t})\,e^{i(k_{c}n-\omega t)}$ and $\rho=(k_{c}/\omega)J_{n}$, and Eq.~\eqref{eq:potwaves} in the conductor becomes
\begin{equation}
\begin{gathered}
\big(k_{c}^{2}-\omega^{2}\mu_{0}\eps\big)\Phi_{t}=\frac{i\sigma_{c}k_{c}}{\eps\,\omega}\big(\omega A_{t}-k_{c}\Phi_{t}\big),\\[2pt]
\big(k_{c}^{2}-\omega^{2}\mu_{0}\eps\big)A_{t}=i\mu_{0}\sigma_{c}\big(\omega A_{t}-k_{c}\Phi_{t}\big).
\end{gathered}
\label{eq:conductorsystem}
\end{equation}
Multiplying the second equation by $\omega$, the first by $k_{c}$, and subtracting,
\begin{equation}
\big(k_{c}^{2}-\omega^{2}\mu_{0}\eps\big)\Big(1+\frac{i\sigma_{c}}{\eps\,\omega}\Big)\big(\omega A_{t}-k_{c}\Phi_{t}\big)=0,
\label{eq:conductorroot}
\end{equation}
so for $\sigma_{c}\neq0$ either $\omega A_{t}=k_{c}\Phi_{t}$, after which Eq.~\eqref{eq:conductorsystem} leaves $k_{c}^{2}=\omega^{2}\mu_{0}\eps$, or $k_{c}^{2}=\omega^{2}\mu_{0}\eps$, after which Eq.~\eqref{eq:conductorsystem} leaves $\omega A_{t}=k_{c}\Phi_{t}$. In both cases
\begin{equation}
\begin{gathered}
k_{c}=\frac{\omega}{c},\qquad A_{t}=\frac{\Phi_{t}}{c},\qquad J_{n}=0,\qquad \rho=0,\qquad E_{n}=0,\\[2pt]
C=ik_{c}A_{t}-\frac{ik_{c}}{c}\Phi_{t}=0\qquad(n>0).
\end{gathered}
\label{eq:gaugewave}
\end{equation}
For plane-wave potentials with constant amplitudes, the only longitudinal solution in the conductor is a wave of pure potentials, with no field, no current, no charge and no scalar, which is the result of Section~2.4 of Ref.~\cite{minotti2023}. The two solutions are matched at $n=0$ by continuity of $\Phi$ and $A_{n}$, which keeps the fields finite, and by the surface charge, $\eps[E_{n}(0^{+})-E_{n}(0^{-})]=\sigma_{f,s}$ with $\partial\sigma_{f,s}/\partial t$ supplied by the current arriving from the conductor. Since that current vanishes, $\sigma_{f,s}=0$ and $E_{n}$ is continuous. With $E_{n}=-\partial\Phi/\partial n-\partial A_{n}/\partial t$ the three conditions read
\begin{equation}
\Phi_{i}+\Phi_{r}=\Phi_{t},\qquad A_{i}+A_{r}=A_{t},\qquad -ik\,(\Phi_{i}-\Phi_{r})+i\omega\,(A_{i}+A_{r})=0,
\label{eq:matchconditions}
\end{equation}
and with $A_{t}=\Phi_{t}/c$ from Eq.~\eqref{eq:gaugewave} and $\omega=ck$ they give
\begin{equation}
\Phi_{r}=0,\qquad \Phi_{t}=\Phi_{i},\qquad A_{t}=\frac{\Phi_{i}}{c},\qquad A_{r}=\frac{\Phi_{i}}{c}-A_{i}.
\label{eq:matchresult}
\end{equation}
The scalar of each wave follows from Eq.~\eqref{eq:Cdef}, with $\partial/\partial n\to\pm ik$ and $\partial/\partial t\to-i\omega$,
\begin{equation}
\begin{gathered}
C_{i}=ik\Big(A_{i}-\frac{\Phi_{i}}{c}\Big),\qquad
C_{t}=ik\Big(A_{t}-\frac{\Phi_{t}}{c}\Big)=0,\\[2pt]
C_{r}=-ik\Big(A_{r}+\frac{\Phi_{r}}{c}\Big)=ik\Big(A_{i}-\frac{\Phi_{i}}{c}\Big)=C_{i},
\end{gathered}
\label{eq:matchscalar}
\end{equation}
so that on the two sides of the wall
\begin{equation}
\begin{gathered}
C=2C_{i}\cos(kn)\,e^{-i\omega t}\quad(n<0),\qquad C=0\quad(n>0),\\[2pt]
C\big|_{0^{-}}=2C_{i}\,e^{-i\omega t},\qquad \frac{\partial C}{\partial n}\Big|_{0^{-}}=0.
\end{gathered}
\label{eq:physicalcondition}
\end{equation}
The condition at a physical conductor is therefore $\partial C/\partial n=0$ on the vacuum side, reached without any assumption about $C$ at the wall, and $C$ itself is discontinuous across the surface. At perpendicular incidence the matching introduces no surface charge and no sheet current. At oblique incidence the same matching gives Eq.~(21) of Ref.~\cite{minotti2023}, in which the reflected scalar is accompanied by a transverse wave, and Eq.~\eqref{eq:reflections} below reproduces it in field form. The same condition also follows from the field equations of this paper. On the vacuum side of a plane wall, with $\sigma_{f,s}=-\eps E_{n}$ and $\vK_{f,s}=\nhat_{c}\times\vB/\mu_{0}$, the surface divergence of the sheet current is $\nabla_{s}\!\cdot\vK_{f,s}=\nhat\cdot(\curl\vB)/\mu_{0}$, and the component of Eq.~\eqref{eq:eedfields} perpendicular to the wall gives
\begin{equation}
\frac{\partial\sigma_{f,s}}{\partial t}+\nabla_{s}\!\cdot\vK_{f,s}
=-\eps\frac{\partial E_{n}}{\partial t}+\frac{1}{\mu_{0}}\Big(\mu_{0}\eps\frac{\partial E_{n}}{\partial t}+\frac{\partial C}{\partial n}\Big)
=\frac{1}{\mu_{0}}\frac{\partial C}{\partial n}.
\label{eq:wallbalance}
\end{equation}
With no current across the surface, Eq.~\eqref{eq:interfacecont} sets the left side to zero, so a conductor that conserves its surface charge and carries no current into its surface has $\partial C/\partial n=0$ on the vacuum side, in agreement with Eq.~\eqref{eq:physicalcondition}.
\\

The difference between the two boundary conditions can be seen directly in reflection. Consider a scalar plane wave incident at an angle $\theta$ to the direction perpendicular to the wall, with $0\leq\theta<\pi/2$,
\begin{equation}
C=C_{i}\,e^{i(\mathbf{k}\cdot\vr-\omega t)},\qquad
|\mathbf{k}|=\frac{\omega}{c},\qquad
\vE=cC\,\frac{\mathbf{k}}{|\mathbf{k}|},\qquad
\vB=\mathbf0.
\label{eq:incident}
\end{equation}
Here $k=|\mathbf{k}|=\omega/c$ is the free-space wavenumber for this calculation, rather than the propagation constant along the axis used elsewhere. Write $C_{r}$ for the reflected scalar amplitude and $B_{r}$ for the signed magnetic amplitude of the reflected transverse wave, whose magnetic field is perpendicular to the plane of incidence. Every term carries the same factor along the wall and in time, which divides out, so the fields on the vacuum side at the wall satisfy
\begin{equation}
\begin{gathered}
C\big|_{\mathrm{wall}}=C_{i}+C_{r},\qquad
\frac{\partial C}{\partial n}\Big|_{\mathrm{wall}}
=ik\cos\theta\,(C_{i}-C_{r}),\\[2pt]
E_{\parallel}\big|_{\mathrm{wall}}
=c\sin\theta\,(C_{i}+C_{r})-c\cos\theta\,B_{r}.
\end{gathered}
\label{eq:wallvalues}
\end{equation}
The conducting wall requires $E_{\parallel}=0$. Combining this with each scalar boundary condition gives
\begin{equation}
\begin{aligned}
C_{r}&=-C_{i}, &
B_{r}&=0, &
C\big|_{\mathrm{wall}}&=0 &
&\text{for } C=0,\\
C_{r}&=C_{i}, &
B_{r}&=2C_{i}\tan\theta, &
C\big|_{\mathrm{wall}}&=2C_{i} &
&\text{for } \partial C/\partial n=0.
\end{aligned}
\label{eq:reflections}
\end{equation}
These wall values are taken on the vacuum side, with $C=0$ inside the conductor. Under the Dirichlet condition, reflection remains purely scalar at every non-grazing angle. The SL modes are therefore not restricted to waves incident perpendicular to the wall. In the rectangular guide of Section~\ref{sec:geom}, pairs of waves reflected in this way form the transverse standing patterns. Under $\partial C/\partial n=0$, the amplitudes instead reproduce Eqs.~(21c) and~(21d) of Ref.~\cite{minotti2023} in field form, while only the pure-potential wave of its Eq.~(21b) enters the conductor. Wherever $C$ varies along the wall, a classical EM component is needed to cancel the parallel electric field of the scalar component. At perpendicular incidence, $\theta=0$, no transverse wave is generated under either condition.
\\

This difference also determines whether a scalar mode can propagate independently in a uniform source-free guide. If $\partial C/\partial n=0$ holds on every side wall, a purely scalar mode with nonzero propagation constant along the axis would also require $C=0$ there by Proposition~\ref{prop:bc}. A solution of the transverse Helmholtz equation whose value and perpendicular derivative both vanish on the wall is identically zero, by the uniqueness theorem for the Cauchy problem of an elliptic equation with analytic coefficients~\cite{couranthilbert2}. Under this alternative condition, every propagating mode with a nonzero scalar component must therefore also contain a classical EM component. Those coupled modes are the modes of a guide bounded by the conductor of Ref.~\cite{minotti2023}, and their spectrum is a separate problem.
\\

The condition $C=0$ of Eq.~\eqref{eq:Cwallchoice} is therefore not the condition obtained by matching at a conductor described by its constitutive laws. It is the condition under which the scalar and Maxwell sectors separate, Section~\ref{sec:slsector}, and the condition imposed on the cleaning scalar at a perfect conductor in hyperbolic divergence cleaning~\cite{munz2000,munzomnesschneider,crestetto2012}. The spectra, mode catalogues and benchmarks that follow are results for that problem. The wall charge that its modes require, and what would have to supply it, are examined in Section~\ref{sec:wallcharge}. The coupled modes of a guide bounded by the conductor of Ref.~\cite{minotti2023} are not treated here.
\\

The wall condition at a physical conductor is obtained by matching the solutions on the two sides of the surface. One of the authors (F.M.) has proposed a realization of that matching in which a thin layer inside the conductor carries a current between two surface charge densities of opposite sign, with the thickness of the layer set by the conductivity and the incident potentials. The constitutive relation that holds within such a layer, Ohm's law or a microscopic relation of the Drude type, and the completion of the matched solution across the surface are left to a separate paper. The results of the present paper do not depend on them, since they are those of the condition $C=0$.
\\

Let $\nhat_c=-\nhat$ point from the conductor into the guide. The ideal-conductor wall sources are
\begin{equation}
\sigma_{f,s}=\eps\,\vE\cdot\nhat_c,
\qquad
\vK_{f,s}=\frac{1}{\mu_{0}}\nhat_c\times\vB.
\label{eq:wallsources}
\end{equation}

\section{Scalar-Longitudinal Sector in the Closed Guide}
\label{sec:slsector}
\subsection{Reduction}
For $\kc\ne0$, the transverse fields again follow from the axial fields. Carrying the $\gradt C$ term from the extended Amp\`ere's law gives
\begin{align}
\Et&=\frac{i}{\kc^{2}}\left[k\,\gradt\Ez-\omega\,\zhat\times\gradt\Bz-\omega\,\gradt C\right],
\label{eq:trans}\\
\Bt&=\frac{i}{\kc^{2}}\left[k\,\gradt\Bz+\omega\mu_{0}\eps\,\zhat\times\gradt\Ez-k\,\zhat\times\gradt C\right].
\label{eq:transB}
\end{align}
Equations~\eqref{eq:trans} and~\eqref{eq:transB} are similar to the classical relations~\eqref{eq:transM} and~\eqref{eq:transBM}, but with two added terms in $\gradt C$, and setting $C=0$ recovers them. Setting $\Bz=0$ and $\Ez=(k/(\omega\mu_{0}\eps))C$ instead gives
\begin{equation}
\begin{aligned}
\Et&=-\frac{i}{\omega\mu_{0}\eps}\,\gradt C,\\[2pt]
\Bt&=\mathbf0,
\end{aligned}
\label{eq:transSL}
\end{equation}
because the two $\gradt C$ terms in $\Bt$ cancel. Equation~\eqref{eq:transSL} is the scalar-longitudinal sector obtained directly from the general reduction.
The classical sectors carry the wall data of Section~\ref{sec:classical}.
\\
 
The scalar sector follows by setting $\vB=\mathbf0$. For $\omega\ne0$, the extended Amp\`ere's law in Eq.~\eqref{eq:eedfields} gives

\begin{equation}
\begin{aligned}
\vE&=-\frac{i}{\omega\mu_{0}\eps}\grad C,\\[2pt]
\Ez&=\frac{k}{\omega\mu_{0}\eps}C,\\[2pt]
\Et&=-\frac{i}{\omega\mu_{0}\eps}\gradt C.
\end{aligned}
\label{eq:slfields}
\end{equation}
\medskip

\noindent The term ``scalar-longitudinal'' means that $C$ generates the axial field $\Ez=(k/(\omega\mu_{0}\eps))C$ as well as $\Et$. It does not mean that the field is purely axial. Because $\vE$ is a gradient, it is irrotational and satisfies Faraday's law identically. The modified Gauss's law becomes

\begin{equation}
(\lapt+\kc^{2})C=0\quad\text{in }\cA.
\label{eq:scalarhelmholtz}
\end{equation}
\medskip

Equation~\eqref{eq:slfields} solves the full extended system, not only the scalar Helmholtz equation. With $\vB=\mathbf0$ and $\vE$ a gradient, Faraday's law and $\dive\vB=0$ hold automatically. In the source-free guide, the extended Amp\`ere's law reduces to $\partial\vE/\partial t=-\frac{\grad C}{\mu_{0}\eps}$, which is satisfied by $\partial/\partial t\to-i\omega$ and $\vE=-(i/(\omega\mu_{0}\eps))\grad C$. The modified Gauss's law then gives Eq.~\eqref{eq:scalarhelmholtz}. Therefore, $(\vE,C)$ with $\vB=\mathbf0$ is a homogeneous source-free solution inside the guide. The wall traces of the adopted exterior problem are considered next.
 
\subsection{Parallel and Perpendicular Wall Conditions}
On the side wall, $\nhat\times\vE=0$, with $\nhat$ the outward transverse normal. Since both $\zhat$ and the boundary tangent lie along the wall, the axial and in-plane parallel electric fields must vanish there.
\\
 
\begin{proposition}[Wall Condition on a Propagating SL Mode]
\label{prop:bc}
For an SL mode with $\omega\ne0$ and $k\ne0$, the perfect-conductor condition $\vE_{\parallel}=\mathbf0$ on the wall is equivalent to
\begin{equation}
C|_{\bnd}=0.
\label{eq:Cdirichlet}
\end{equation}

At cutoff, $k=0$, the axial field vanishes identically and the perfect-conductor condition no longer fixes $C$ on the wall. There Eq.~\eqref{eq:Cdirichlet} is adopted as the limit of the propagating mode.
\end{proposition}
 
\begin{proof}
For $k\ne0$, Eq.~\eqref{eq:slfields} gives $\Ez=(k/(\omega\mu_{0}\eps))C$. Therefore $\Ez=0$ on the side wall implies $C=0$ on $\bnd$. Similarly, if $C=0$ on the boundary curve, its derivative along the wall also vanishes on smooth segments.  Therefore, the wall-parallel part of $\Et=-(i/(\omega\mu_{0}\eps))\gradt C$ vanishes, as does $\Ez$. At cutoff, $\Ez$ alone no longer fixes $C$. The homogeneous Dirichlet condition is then taken as the limit of the propagating mode and the homogeneous Dirichlet scalar wall condition.
\end{proof}
 
At cutoff the conductor fixes only the derivative of $C$ along the wall, so $C=0$ is imposed there as the limit of the propagating mode. For the full time-dependent problem the condition is an independent wall condition, as Section~\ref{sec:eed} explained.
 
\subsection{Cavity Limit}
Closing the guide with perfectly conducting end walls at $z=0$ and $z=L$ adds the boundary $C=0$ there as well, since $\vE_{\parallel}=\mathbf0$ on an end wall makes $\gradt C$ vanish on it by Eq.~\eqref{eq:slfields}, so $C$ is constant on the end wall and equal to its side-wall value zero. The axial dependence of a separated SL mode then becomes a standing wave,

\begin{equation}
C=A\,\phi_j(x,y)\sin\left(\frac{p\pi z}{L}\right)\,e^{-i\omega t},
\qquad
\omega^{2}\mu_{0}\eps=\lambda_j+\left(\frac{p\pi}{L}\right)^{2},
\qquad
p=1,2,\dots,
\label{eq:cavitylimit}
\end{equation}
where $\phi_j$ and $\lambda_j$ are the Dirichlet eigenpairs of Section~\ref{sec:theorem}. Each transverse SL family of the guide gives resonator modes at the allowed axial wavenumbers $k=p\pi/L$ with $p\geq1$, while a traveling guide mode does not survive at an axial wavenumber outside that set. The cutoff member does not survive either, since $p=0$ gives $C\equiv0$, and the SL spectrum of the resulting right cylinder is the TM spectrum with axial index $p\ge1$, mode for mode. That is the resonator counterpart of the theorem below and is developed in another paper for closed cavities.
 
\subsection{Wall Charge of an SL Mode}
\label{sec:wallcharge}
As in a TM mode, the perpendicular electric field of an SL mode ends on induced wall charge. With $\nhat_c=-\nhat$ pointing from the conductor into the guide, Eq.~\eqref{eq:wallsources} gives $\sigma_{f,s}=\eps\,\vE\cdot\nhat_c$ and $\vK_{f,s}=\nhat_c\times\vB/\mu_{0}$, and for an SL mode $\vB=\mathbf0$, so $\vK_{f,s}=\mathbf0$ while $\sigma_{f,s}$ oscillates. On the wall $x=0$ of the rectangular SL$_{nm}$ mode, $\nhat_c=\mathbf e_x$ and Eq.~\eqref{eq:rectEx} gives

\begin{equation}
\sigma_{f,s}(0,y)=\eps E_x(0,y)
=-\frac{i}{\mu_{0}\omega}\frac{n\pi}{a}\,C_0\sin\left(\frac{m\pi y}{b}\right),
\label{eq:rectsigma}
\end{equation}
\medskip

\noindent and the other walls follow in the same way. Charge conservation on the interface, with $\vJ_{f,g}$ and $\vJ_{f,c}$ the free current densities on the guide and conductor sides and $\nabla_s\!\cdot$ the surface divergence, reads
\begin{equation}
\frac{\partial\sigma_{f,s}}{\partial t}+\nabla_s\!\cdot\vK_{f,s}=\nhat_c\cdot\left(\vJ_{f,c}-\vJ_{f,g}\right).
\label{eq:interfacecont}
\end{equation}
and in the source-free guide $\vJ_{f,g}=\mathbf0$ and $\vK_{f,s}=\mathbf0$. With $\sigma_{f,s}=\eps\,\vE\cdot\nhat_{c}$ on the guide side and Eq.~\eqref{eq:slfields}, an SL mode has
\begin{equation}
\sigma_{f,s}=-\frac{i}{\omega\mu_{0}}\,\grad C\cdot\nhat_{c}=\frac{i}{\omega\mu_{0}}\frac{\partial C}{\partial n},\qquad \pdt{\sigma_{f,s}}=-i\omega\,\sigma_{f,s}=\frac{1}{\mu_{0}}\frac{\partial C}{\partial n},
\label{eq:slwallcharge}
\end{equation}
and no parallel current feeds this charge. It is accounted for in one of two ways,
\begin{equation}
\begin{aligned}
\nhat_{c}\cdot\vJ_{f,c}&=\frac{1}{\mu_{0}}\frac{\partial C}{\partial n}, & \mu_{0}\vJ_{f,c}&=-\grad C & &\text{charge conserved at the wall},\\
\Lambda_{s}&=\frac{1}{\mu_{0}}\frac{\partial C}{\partial n}, & C&=0 & &\text{no current in the conductor},
\end{aligned}
\label{eq:wallanomaly}
\end{equation}
with the second holding in the conductor. The first line uses $\vE=\vB=\mathbf0$ in the conductor at nonzero frequency and needs a scalar gradient there, outside the vacuum problem. The second holds when $C=0$ in the conductor, as Section~2.4 of Ref.~\cite{minotti2023} finds under its material assumptions for a conductor with locally conserved current, and is the singular part of Eq.~\eqref{eq:Cwave} for a scalar that ends at the wall. The interior of the conductor is not modeled. At the conductor of Ref.~\cite{minotti2023}, which conserves its surface charge and carries no current into its surface, neither reading applies. Equation~\eqref{eq:interfacecont} then gives $\partial C/\partial n=0$ on the vacuum side rather than $C=0$, Eq.~\eqref{eq:wallbalance}, and the SL modes of this paper are not modes of a guide bounded by that conductor. The SL field therefore ends on a time-harmonic wall charge with no current parallel to the wall,

\begin{equation}
\vK_{f,s}=\mathbf 0
\qquad\text{for every SL mode,}
\label{eq:noKs}
\end{equation}

\vspace{5pt}

\noindent which is the wall-side signature of $\vB=\mathbf0$. TE and TM modes carry parallel wall current and may carry charge as well, TE$_{10}$ on the walls $y=0,b$ for instance, and the SL family is the only one with charge and no sheet current.

\subsection{Maxwell-SL Decomposition}
\begin{proposition}[Maxwell and SL Parts of a Field]
\label{prop:decomp}
At a fixed frequency $\omega\ne0$, every source-free field with $C=0$ on the wall splits uniquely into a Maxwell field and an SL field,

\begin{equation}
(\vE,\vB,C)=(\vE_{M},\vB_{M},0)+(\vE_{\mathrm{SL}},\mathbf0,C),
\qquad
\vE_{\mathrm{SL}}=-\frac{i}{\omega\mu_{0}\eps}\grad C,
\end{equation}
\medskip

\noindent with $(\vE_{M},\vB_{M})$ obeying Maxwell's equations and the perfect-conductor conditions.
\end{proposition}

\begin{proof}
$C$ obeys $\nabla^{2}C+\omega^{2}\mu_{0}\eps\,C=0$ by Eq.~\eqref{eq:Cwave}, so $(\vE_{\mathrm{SL}},\mathbf0,C)$ satisfies Eq.~\eqref{eq:eedfields} by substitution. What remains after subtracting it has $C=0$ and is a Maxwell field, and it keeps $\vE_{\parallel}=\mathbf0$ on the wall because $C=0$ there makes $\grad C$, and with it $\vE_{\mathrm{SL}}$, perpendicular to the wall. The separate wall conditions therefore follow from $C=0$ and need no further assumption.
\end{proof}

This separation relies on the boundary condition $C=0$ at the wall. With $\partial C/\partial n=0$, the SL fields still satisfy the field equations on their own. However, wherever $C$ varies along the wall, the SL electric field has a component parallel to the surface. The Maxwell electric field must cancel that component to satisfy the conducting boundary condition. The wall therefore couples the two parts, as the oblique reflection example in Section~\ref{sec:extwall} shows.
\\

If $C$ is uniform along the wall, this coupling is absent. For example, when a scalar wave strikes a plane wall perpendicularly, the amplitude $2|C_{i}|\tan\theta$ is zero. The wave then reflects without generating an ordinary electromagnetic wave under either boundary condition.

\section{Guide Spectral Theorem}
\label{sec:theorem}
\subsection{Dirichlet Eigenproblem}
 
Let $D=-\lapt$ act on functions that vanish on $S$, so that $D\psi=\gamma^{2}\psi$ is the transverse Dirichlet eigenvalue problem $\lapt\psi+\gamma^{2}\psi=0$ with $\psi|_{S}=0$. On a bounded cross-section with piecewise smooth boundary this problem has a discrete sequence of positive eigenvalues of finite multiplicity, with eigenfunctions orthogonal and complete in the cross-section~\cite{mclean2000}.  Denote its eigenpairs by

\begin{equation}
0<\lambda_1\le\lambda_2\le\cdots,
\qquad
D\phi_j=\lambda_j\phi_j,
\qquad
\int_{\cA}\phi_j^{*}\phi_k\,dA=\delta_{jk},
\label{eq:dirichleteigs}
\end{equation}
\medskip

\noindent where repeated eigenvalues are counted with multiplicity. Within a degenerate eigenspace, any orthonormal basis may be used, such as the paired $\cos m\varphi$ and $\sin m\varphi$ modes of the coaxial guide in Section~\ref{sec:geom}.
\\
 
\begin{theorem}[SL Cutoff Spectrum]
\label{thm:slspectrum}
Let $\omega\ne0$. With the homogeneous scalar condition $C|_{\bnd}=0$, every separated exponential SL mode, including every bounded cutoff mode, comes from a Dirichlet eigenpair $(\lambda_j,\phi_j)$ as

\begin{align}
C_j(x,y,z,t)&=A_j\phi_j(x,y)e^{i(k_j z-\omega t)},\label{eq:Cmode}\\[14pt]
\vB_j&=\mathbf0,\\[10pt]
\vE_j&=-\frac{i}{\omega\mu_{0}\eps}\grad C_j.
\end{align}

The propagation constant satisfies

\begin{equation}
\boxed{k_j^2=\omega^{2}\mu_{0}\eps-\lambda_j}.
\label{eq:kj}
\end{equation}
\medskip

The SL cutoff frequencies are therefore

\begin{equation}
\omega_{c,j}^{\mathrm{SL}}=c\sqrt{\lambda_j}.
\end{equation}
\medskip

These are exactly the classical TM cutoff frequencies of the same guide, with the same eigenspace dimension at each Dirichlet eigenvalue. Both sectors solve the same self-adjoint Dirichlet problem. No cutoff-free SL mode exists, so there is no scalar counterpart of the TEM family.
\end{theorem}
 
\begin{proof}
Equations~\eqref{eq:scalarhelmholtz} and~\eqref{eq:Cdirichlet} are the eigenvalue problem for $D$, with $\kc^2=\lambda_j$. This gives Eqs.~\eqref{eq:Cmode} to~\eqref{eq:kj}. Conversely, any Dirichlet eigenfunction inserted into Eq.~\eqref{eq:slfields} satisfies the source-free extended system and the scalar wall data. The TM sector obeys the same problem. Set $C=0$, $\Bz=0$, and $\Ez\not\equiv0$. Eliminating the transverse fields from the Maxwell curl equations gives $\vE_t=(ik/\kc^{2})\gradt\Ez$ and $(\lapt+\kc^{2})\Ez=0$, where $\kc^{2}=\omega^{2}\mu_{0}\eps-k^{2}$. Because the axial direction lies along the side wall, $\nhat\times\vE=0$ gives $\Ez|_{\bnd}=0$. Thus $-\lapt\Ez=\kc^{2}\Ez$ is the same Dirichlet problem as Eq.~\eqref{eq:dirichleteigs}. The TM and SL cutoff wavenumbers are therefore identical. A cutoff-free SL mode would require $\lambda_j=0$, but the first Dirichlet eigenvalue on a bounded cross-section is strictly positive. Equivalently, the only harmonic function that vanishes on all of $\bnd$ is zero.
\end{proof}
 
\begin{corollary}[Scalar Completeness and Conditional Modal Decomposition]
\label{cor:complete}
The Dirichlet eigenfunctions $\phi_j$ of Eq.~\eqref{eq:dirichleteigs} are orthogonal and complete on the cross-section~\cite{couranthilbert}. With the normalization $\int_{\cA}\phi_j^{*}\phi_k\,dA=\delta_{jk}$, any scalar field profile in the guide expands as

\begin{equation}
C(x,y)=\sum_j a_j\phi_j(x,y),
\qquad
a_j=\int_{\cA}\phi_j^{*}C\,dA,
\label{eq:scalarexp}
\end{equation}
exactly as a TM axial field expands in the TM modes. The gradients are orthogonal as well, Eq.~\eqref{eq:gradorth} below, so the Dirichlet integral of a field is the sum of those of its modes, $\int_{\cA}|\gradt C|^{2}\,dA=\sum_j\lambda_j|a_j|^{2}$. Together with Proposition~\ref{prop:decomp}, the TE, TM, TEM and SL modes therefore give the complete modal decomposition of a source-free field at one frequency, each profile carrying both axial branches $e^{\pm ik_jz}$ and, at cutoff, the axial dependence $a_j+b_jz$, where on an infinite guide boundedness sets the linear coefficient to zero, the classical part being the standard waveguide-mode expansion~\cite{hallamonk}.
\end{corollary}
\medskip
 
Green's first identity, whose boundary term vanishes by the Dirichlet data, also gives
\medskip

\begin{equation}
\int_{\cA}\gradt\phi_j^{*}\cdot\gradt\phi_k\,dA=-\int_{\cA}\phi_k\,\lapt\phi_j^{*}\,dA=\lambda_j\delta_{jk}.
\label{eq:gradorth}
\end{equation}
\medskip

These relations hold on any bounded cross-section with piecewise smooth boundary and do not require separation of variables. The same operator also gives a variational description of the cutoffs and supports the numerical work in Section~\ref{sec:lshape}.
 
\section{Canonical Guides}
\label{sec:geom}

The rectangular, circular and coaxial guides are the three standard geometries of Ref.~\cite{jackson}, and each is treated in the same order. The L-shaped guide is added as a fourth example, because it is not separable and has a reentrant corner, so it tests the wall condition on $C$, and the identity of the SL and TM cutoffs, where the full spectrum is not available in closed form. In the source-free interior, $\rho_{\mathrm{tot}}=0$ and $\vJ_{\mathrm{tot}}=\mathbf0$, the field-form wave equations \eqref{eq:fieldwave} and the scalar wave equation \eqref{eq:Cwave} are homogeneous, and for fields with the time dependence $e^{-i\omega t}$ each becomes a Helmholtz equation,
\begin{equation}
\nabla^{2}\vE+\omega^{2}\mu_{0}\eps\,\vE=\mathbf0,
\qquad
\nabla^{2}\vB+\omega^{2}\mu_{0}\eps\,\vB=\mathbf0,
\qquad
\nabla^{2}C+\omega^{2}\mu_{0}\eps\,C=0.
\label{eq:helmholtz3}
\end{equation}

The first two are the classical equations. The third is the extended one, and it has the same form because $C$ obeys the same wave operator. For a guide uniform along $z$ the Laplacian separates,

\begin{equation}
\nabla^{2}=\lapt+\frac{\partial^{2}}{\partial z^{2}},
\qquad
\frac{\partial^{2}}{\partial z^{2}}\to-k^{2}
\quad\text{for the dependence } e^{ikz},
\label{eq:lapsplit}
\end{equation}
\medskip

\noindent so every axial generator $\psi(x,y)$, whether $\Ez$, $\Bz$ or $C$, satisfies the transverse Helmholtz equation

\begin{equation}
\lapt\psi+\kc^{2}\psi=0,
\qquad
\kc^{2}=\omega^{2}\mu_{0}\eps-k^{2},
\label{eq:genhelm}
\end{equation}
\medskip

with $k^{2}>0$ propagating, $k=0$ at cutoff, and $k^{2}<0$ evanescent. The transverse field components follow from the generators through Eqs.~\eqref{eq:transM} and~\eqref{eq:transBM} for the Maxwell sectors and Eq.~\eqref{eq:slfields} for the scalar sector. The wall conditions are fixed before any equation is solved. By Eq.~\eqref{eq:classicaldata}, TM requires $\Ez=0$ on $S$, TE requires $\partial\Bz/\partial n=0$ on $S$, and a TEM field has $\kc=0$ with $\Et=-\gradt\Phi$, $\lapt\Phi=0$, and $\Phi$ constant on each conductor. By Eq.~\eqref{eq:Cdirichlet}, SL requires $C=0$ on $S$. In each geometry Eq.~\eqref{eq:genhelm} is separated, the resulting ordinary differential equations are solved in general, and the four families are then obtained one by one from their wall conditions.

\subsection{Rectangular Guide}
\label{sec:rect}

For $0\le x\le a$ and $0\le y\le b$, the trial solution $\psi=X(x)Y(y)$ in Eq.~\eqref{eq:genhelm} gives $X''Y+XY''+\kc^{2}XY=0$, and dividing by $XY$ separates the variables,

\begin{equation}
X''+k_x^{2}X=0,
\qquad
Y''+k_y^{2}Y=0,
\qquad
k_x^{2}+k_y^{2}=\kc^{2},
\label{eq:rectode}
\end{equation}
\smallskip

\noindent with $k_x^{2}$ and $k_y^{2}$ the separation constants. Each equation is solved by a sine and a cosine, so the general solution is the sum over all separation pairs with independent coefficients,

\begin{equation}
\begin{aligned}
\psi(x,y)=\sum_{n,m}\big(&A_{nm}\sin(k_{x}x)\sin(k_{y}y)+B_{nm}\sin(k_{x}x)\cos(k_{y}y)\\[2pt]
{}+{}&C_{nm}\cos(k_{x}x)\sin(k_{y}y)+D_{nm}\cos(k_{x}x)\cos(k_{y}y)\big),
\end{aligned}
\label{eq:rectgen}
\end{equation}
\smallskip

\noindent with $k_x^{2}+k_y^{2}=\kc^{2}$ in every term. The wall conditions now select the terms and quantize $k_x$ and $k_y$ for each family.

\subsubsection{TM Modes}
The generator field function is $\psi=\Ez$ with $\Ez=0$ on all four walls. The conditions at $x=0$ and $y=0$ remove every term of Eq.~\eqref{eq:rectgen} containing a cosine, and the conditions at $x=a$ and $y=b$ then require $\sin(k_x a)=0$ and $\sin(k_y b)=0$, so

\begin{equation}
k_x=\frac{n\pi}{a},
\qquad
k_y=\frac{m\pi}{b},
\qquad
\Ez=E_0\sin\left(\frac{n\pi x}{a}\right)\sin\left(\frac{m\pi y}{b}\right),
\qquad
n,m\ge1,
\label{eq:recttmgen}
\end{equation}
where $n=0$ or $m=0$ would give $\Ez\equiv0$. Equations~\eqref{eq:transM} and~\eqref{eq:transBM} with $\Bz=0$ and $C=0$ give the complete TM$_{nm}$ field, with the factor $e^{i(kz-\omega t)}$ understood and agreeing with classical physics~\cite{pozar,collin,marcuvitz},

\begin{equation}
\begin{aligned}
E_x&=\frac{ik}{\kc^{2}}\frac{n\pi}{a}E_0\cos\left(\frac{n\pi x}{a}\right)\sin\left(\frac{m\pi y}{b}\right),\\[2pt]
E_y&=\frac{ik}{\kc^{2}}\frac{m\pi}{b}E_0\sin\left(\frac{n\pi x}{a}\right)\cos\left(\frac{m\pi y}{b}\right),\\[2pt]
E_z&=E_0\sin\left(\frac{n\pi x}{a}\right)\sin\left(\frac{m\pi y}{b}\right),\\[2pt]
B_x&=-\frac{i\omega\mu_{0}\eps}{\kc^{2}}\frac{m\pi}{b}E_0\sin\left(\frac{n\pi x}{a}\right)\cos\left(\frac{m\pi y}{b}\right),\\[2pt]
B_y&=\frac{i\omega\mu_{0}\eps}{\kc^{2}}\frac{n\pi}{a}E_0\cos\left(\frac{n\pi x}{a}\right)\sin\left(\frac{m\pi y}{b}\right),\\[2pt]
B_z&=0.
\end{aligned}
\label{eq:catTM}
\end{equation}
\smallskip

The cutoff eigenvalue and the dispersion relation follow from Eq.~\eqref{eq:rectode} and Eq.~\eqref{eq:genhelm},

\begin{equation}
\kc^{2}=\left(\frac{n\pi}{a}\right)^{2}+\left(\frac{m\pi}{b}\right)^{2},
\qquad
k^{2}=\omega^{2}\mu_{0}\eps-\left(\frac{n\pi}{a}\right)^{2}-\left(\frac{m\pi}{b}\right)^{2},
\label{eq:rectcutoff}
\end{equation}
\smallskip

\noindent so the lowest TM mode is TM$_{11}$. The transverse magnetic field of every TM mode is nonzero,

\begin{equation}
\Bt^{\mathrm{TM}}=\frac{i\omega\mu_{0}\eps}{\kc^{2}}\,\zhat\times\gradt\Ez^{\mathrm{TM}}\ne\mathbf0,
\label{eq:recttm}
\end{equation}
\smallskip

\noindent which is what will distinguish it from the scalar mode with the same profile.

\subsubsection{TE Modes}
The generator is $\psi=\Bz$ with $\partial\Bz/\partial n=0$ on all four walls. The conditions $\partial\Bz/\partial x=0$ at $x=0$ and $\partial\Bz/\partial y=0$ at $y=0$ remove every term of Eq.~\eqref{eq:rectgen} containing a sine, and the same conditions at $x=a$ and $y=b$ quantize the wavenumbers as in Eq.~\eqref{eq:recttmgen}, now with the cosine products,

\begin{equation}
\Bz=B_0\cos\left(\frac{n\pi x}{a}\right)\cos\left(\frac{m\pi y}{b}\right),
\qquad
n,m\ge0,
\qquad
(n,m)\ne(0,0),
\label{eq:recttegen}
\end{equation}
since one index may vanish without killing the field but $(0,0)$ gives a constant $\Bz$, which carries no transverse field. Equations~\eqref{eq:transM} and~\eqref{eq:transBM} with $\Ez=0$ and $C=0$ give the complete TE$_{nm}$ field,

\begin{equation}
\begin{aligned}
E_x&=-\frac{i\omega}{\kc^{2}}\frac{m\pi}{b}B_0\cos\left(\frac{n\pi x}{a}\right)\sin\left(\frac{m\pi y}{b}\right),\\[2pt]
E_y&=\frac{i\omega}{\kc^{2}}\frac{n\pi}{a}B_0\sin\left(\frac{n\pi x}{a}\right)\cos\left(\frac{m\pi y}{b}\right),\\[2pt]
E_z&=0,\\[2pt]
B_x&=-\frac{ik}{\kc^{2}}\frac{n\pi}{a}B_0\sin\left(\frac{n\pi x}{a}\right)\cos\left(\frac{m\pi y}{b}\right),\\[2pt]
B_y&=-\frac{ik}{\kc^{2}}\frac{m\pi}{b}B_0\cos\left(\frac{n\pi x}{a}\right)\sin\left(\frac{m\pi y}{b}\right),\\[2pt]
B_z&=B_0\cos\left(\frac{n\pi x}{a}\right)\cos\left(\frac{m\pi y}{b}\right),
\end{aligned}
\label{eq:catTE}
\end{equation}
\smallskip

\noindent with the same cutoff eigenvalue and dispersion relation as Eq.~\eqref{eq:rectcutoff}. For $a>b$ the lowest mode of the guide is TE$_{10}$ at $\kc=\pi/a$, and when $a=b$ it is degenerate with TE$_{01}$.

\subsubsection{TEM Mode}
A TEM field has $\kc=0$, so Eq.~\eqref{eq:genhelm} reduces to the Laplace equation $\lapt\Phi=0$ for the transverse potential, with $\Phi$ equal to one constant on the single connected conductor. By uniqueness of the Dirichlet problem, $\Phi$ takes that constant value everywhere, so $\Et=-\gradt\Phi=\mathbf0$ and the rectangular guide carries no TEM mode.

\subsubsection{SL Modes}
The generator is now $\psi=C$ with $C=0$ on all four walls in the extended frame, the same data as the TM generator. The same selection therefore keeps only the sine products of Eq.~\eqref{eq:rectgen} and quantizes the wavenumbers exactly as in Eq.~\eqref{eq:recttmgen},

\begin{equation}
\begin{aligned}
C_{nm}(x,y)&=C_0\sin\left(\frac{n\pi x}{a}\right)\sin\left(\frac{m\pi y}{b}\right),\\[6pt]
\kc^2&=\left(\frac{n\pi}{a}\right)^2+\left(\frac{m\pi}{b}\right)^2,\qquad n,m\ge1.
\end{aligned}
\label{eq:rect}
\end{equation}
\smallskip

Equation~\eqref{eq:transSL} gives the Cartesian field components

\begin{align}
E_x&=-\frac{i}{\omega\mu_{0}\eps}\frac{n\pi}{a}\,C_0\cos\left(\frac{n\pi x}{a}\right)\sin\left(\frac{m\pi y}{b}\right),
\label{eq:rectEx}\\
E_y&=-\frac{i}{\omega\mu_{0}\eps}\frac{m\pi}{b}\,C_0\sin\left(\frac{n\pi x}{a}\right)\cos\left(\frac{m\pi y}{b}\right),
\label{eq:rectEy}\\
E_z&=\frac{k}{\omega\mu_{0}\eps}\,C_0\sin\left(\frac{n\pi x}{a}\right)\sin\left(\frac{m\pi y}{b}\right),
\label{eq:rectEz}
\end{align}

with the factor $e^{i(k z-\omega t)}$ understood and $\vB=\mathbf0$. Direct substitution checks the field equations. Faraday's law holds because $\vE$ is a gradient and $\vB=\mathbf0$. The extended Amp\`ere's law becomes $-\mu_{0}\eps\,\partial\vE/\partial t=\gradt C+ik C\zhat$, which Eqs.~\eqref{eq:rectEx} to~\eqref{eq:rectEz} satisfy component by component. The modified Gauss's law gives

\begin{equation}
\dive\vE+\frac{\partial C}{\partial t}
=-\frac{i}{\omega\mu_{0}\eps}\nabla^{2}C-i\omega C
=\frac{i}{\omega\mu_{0}\eps}\left(\kc^{2}+k^{2}\right)C-i\omega C=0
\label{eq:rectgauss}
\end{equation}
\smallskip

\noindent when $\kc^{2}+k^{2}=\omega^{2}\mu_{0}\eps$. On every wall the two electric components parallel to the wall vanish with $C$, while the perpendicular component may remain nonzero. The cutoff eigenvalue is Eq.~\eqref{eq:rectcutoff}, the TM value, so the first SL mode is SL$_{11}$ and shares its cutoff with TM$_{11}$. At a common index $(n,m)$ away from cutoff, Eqs.~\eqref{eq:catTM} and~\eqref{eq:rectEx} to~\eqref{eq:rectEz} give

\begin{equation}
\left.\Et\right|_{\mathrm{SL}}
=-\frac{\kc^{2}}{\omega k\mu_{0}\eps}\frac{C_0}{E_0}\left.\Et\right|_{\mathrm{TM}},
\qquad
\left.\Bt\right|_{\mathrm{SL}}=\mathbf0,
\qquad
\left.\Bt\right|_{\mathrm{TM}}\ne\mathbf0.
\label{eq:catcompare}
\end{equation}
\medskip

The axial components obey a different scaling,

\begin{equation}
\left.\Ez\right|_{\mathrm{SL}}
=\frac{k}{\omega\mu_{0}\eps}\frac{C_0}{E_0}\left.\Ez\right|_{\mathrm{TM}},
\qquad
\frac{k/\omega}{-\kc^{2}/(\omega k)}=-\frac{k^{2}}{\kc^{2}}\ne1,
\label{eq:catEzcompare}
\end{equation}
\medskip

where the ratio on the right also holds for evanescent modes and is never one at nonzero frequency. A single amplitude rescaling cannot map the full TM electric field to the SL field. Only their transverse patterns are proportional. The magnetic field distinguishes them immediately, since it is present only in TM, while $C$ is present only in SL. At exact cutoff, $k=0$ and $\omega=c\kc$, so

\begin{equation}
\left.\Et\right|_{\mathrm{TM}}=\mathbf0,
\qquad
\left.\Ez\right|_{\mathrm{SL}}=0,
\qquad
\left.\Ez\right|_{\mathrm{TM}}\ne0,
\qquad
\left.\Et\right|_{\mathrm{SL}}\ne\mathbf0,
\label{eq:catcutoff}
\end{equation}
\medskip

so the cutoff fields are disjoint. The TM electric field is purely axial, while the SL electric field is purely transverse. Corollary~\ref{cor:complete} expands any scalar field in the guide as the double sum

\begin{equation}
\resizebox{0.9\linewidth}{!}{$\displaystyle
\begin{gathered}
C(x,y,z,t)=\sum_{n\ge1}\sum_{m\ge1}\sin\left(\frac{n\pi x}{a}\right)\sin\left(\frac{m\pi y}{b}\right)\left[c^{+}_{nm}e^{ik_{nm}z}+c^{-}_{nm}e^{-ik_{nm}z}\right]e^{-i\omega t},\\
k_{nm}^{2}=\omega^{2}\mu_{0}\eps-\left(\frac{n\pi}{a}\right)^{2}-\left(\frac{m\pi}{b}\right)^{2},
\end{gathered}
$}
\label{eq:rectsum}
\end{equation}
\smallskip

\noindent and the TM and TE families expand over Eqs.~\eqref{eq:catTM} and~\eqref{eq:catTE}, with $n,m\ge1$ for TM and $n,m\ge0$ with $(n,m)\ne(0,0)$ for TE. At cutoff, $k_{nm}=0$, the bracket becomes $c^{+}_{nm}+c^{-}_{nm}z$. On an infinite guide boundedness sets the linear coefficient to zero, which is the part Theorem~\ref{thm:slspectrum} keeps. The junction analysis of Section~\ref{sec:matching} is built on this expansion.

\subsection{Circular Cylinder Guide}
\label{sec:circ}

The cross-section is the disk $0\le r<a$ in cylindrical coordinates $(r,\varphi,z)$, with the wall at $r=a$. In these coordinates Eq.~\eqref{eq:genhelm} reads

\begin{equation}
\frac{1}{r}\frac{\partial}{\partial r}\left(r\frac{\partial\psi}{\partial r}\right)+\frac{1}{r^{2}}\frac{\partial^{2}\psi}{\partial\varphi^{2}}+\kc^{2}\psi=0,
\label{eq:circhelm}
\end{equation}
\smallskip

\noindent and the trial solution $\psi=R(r)e^{in\varphi}$, with $n$ an integer so that $\psi$ is single valued in $\varphi$, separates it into Bessel's equation

\begin{equation}
r^{2}R''+r R'+\left(\kc^{2}r^{2}-n^{2}\right)R=0,
\qquad
R=A\,J_{n}(\kc r)+B\,Y_{n}(\kc r).
\label{eq:circbessel}
\end{equation}
\smallskip

\noindent Finiteness at the origin excludes $Y_{n}$, so $B=0$ and $R=A\,J_{n}(\kc r)$ for all. The wall condition then selects $\kc$.

\subsubsection{TM Modes}
With $\Ez=J_{n}(\kc r)e^{in\varphi}$, the condition $\Ez=0$ at $r=a$ gives

\begin{equation}
J_{n}(\kc a)=0,
\qquad
\kc=\frac{x_{nm}}{a},
\qquad
m=1,2,\dots,
\label{eq:circtm}
\end{equation}
where $x_{nm}$ is the $m$th positive zero of $J_{n}$. Equations~\eqref{eq:transM} and~\eqref{eq:transBM} with $\Bz=0$ give the transverse fields

\begin{equation}
\begin{aligned}
E_{r}&=\frac{ik}{\kc}\,J_{n}'(\kc r)e^{in\varphi},
&\qquad
E_{\varphi}&=-\frac{kn}{\kc^{2}r}\,J_{n}(\kc r)e^{in\varphi},
\\[2pt]
B_{r}&=\frac{\omega\mu_{0}\eps\,n}{\kc^{2}r}\,J_{n}(\kc r)e^{in\varphi},
&\qquad
B_{\varphi}&=\frac{i\omega\mu_{0}\eps}{\kc}\,J_{n}'(\kc r)e^{in\varphi},
\end{aligned}
\label{eq:circtmfields}
\end{equation}
\medskip

\noindent with $C=0$ and the common factor $e^{i(kz-\omega t)}$. The cutoff frequency is $\omega_{c}=x_{nm}/(a\sqrt{\mu_{0}\eps})$ and the dispersion relation is $k^{2}=\omega^{2}\mu_{0}\eps-x_{nm}^{2}/{a^{2}}$. The lowest TM mode is the axisymmetric TM$_{01}$ at $x_{01}=2.405$.

\subsubsection{TE Modes}

With $\Bz=J_{n}(\kc r)e^{in\varphi}$, the Neumann condition $\partial\Bz/\partial r=0$ at $r=a$ gives

\begin{equation}
J_{n}'(\kc a)=0,
\qquad
\kc=\frac{x'_{nm}}{a},
\label{eq:circte}
\end{equation}

\newpage

\noindent with $x'_{nm}$ the $m$th positive zero of $J_{n}'$, and the same relations with $\Ez=0$ give

\begin{equation}
\begin{aligned}
B_{r}&=\frac{ik}{\kc}\,J_{n}'(\kc r)e^{in\varphi},
&\qquad
B_{\varphi}&=-\frac{kn}{\kc^{2}r}\,J_{n}(\kc r)e^{in\varphi},
\\[2pt]
E_{r}&=-\frac{\omega n}{\kc^{2}r}\,J_{n}(\kc r)e^{in\varphi},
&\qquad
E_{\varphi}&=-\frac{i\omega}{\kc}\,J_{n}'(\kc r)e^{in\varphi}.
\end{aligned}
\label{eq:circtefields}
\end{equation}
\medskip

The cutoff is $\omega_{c}=x'_{nm}/(a\sqrt{\mu_{0}\eps})$. The dominant mode of the circular guide is TE$_{11}$ at $x'_{11}=1.841$, and because $J_{0}'=-J_{1}$ every $x'_{0m}$ equals $x_{1m}$, so each TE$_{0m}$ is degenerate with TM$_{1m}$.

\subsubsection{TEM Mode}
With $\kc=0$ the transverse potential obeys $\lapt\Phi=0$ in the disk with $\Phi$ constant on the wall. The disk is simply connected, so by uniqueness $\Phi$ is constant throughout, $\Et=\mathbf0$, and the circular guide carries no TEM mode.

\subsubsection{SL Modes}

With $C=J_{n}(\kc r)e^{in\varphi}$, the condition $C=0$ gives the same wall condition as TM, $J_{n}(\kc a)=0$ and $\kc=x_{nm}/a$, and Eq.~\eqref{eq:slfields} gives

\begin{equation}
\begin{aligned}
E_{r}&=-\frac{i\kc}{\omega\mu_{0}\eps}\,J_{n}'(\kc r)e^{in\varphi},
&\qquad
E_{\varphi}&=\frac{n}{\omega\mu_{0}\eps\,r}\,J_{n}(\kc r)e^{in\varphi},
\\[2pt]
\Ez&=\frac{k}{\omega\mu_{0}\eps}\,J_{n}(\kc r)e^{in\varphi},
&\qquad
\vB&=\mathbf0.
\end{aligned}
\label{eq:circsl}
\end{equation}
\medskip

The electric field is the gradient of $C$ and has no magnetic counterpart, and its radial profile is the Bessel function of the TM mode with the same indices. The cutoff frequencies of all three families are therefore

\begin{equation}
\omega_{c}^{\mathrm{TM}}=\omega_{c}^{\mathrm{SL}}=\frac{x_{nm}}{a\sqrt{\mu_{0}\eps}},
\qquad
\omega_{c}^{\mathrm{TE}}=\frac{x'_{nm}}{a\sqrt{\mu_{0}\eps}},
\label{eq:circcutoff}
\end{equation}
\smallskip

\noindent and Table~\ref{tab:cutoffs} lists the lowest values. Every mode with $n\ge1$ is two-fold degenerate through $e^{\pm in\varphi}$. For the power of an SL mode, Proposition~\ref{prop:powerenergy} uses the radial norm $\int_{0}^{a}J_{n}^{2}(x_{nm}r/a)\,r\,d r=\tfrac12a^{2}J_{n+1}^{2}(x_{nm})$, which follows from Lommel's integral at a zero of $J_{n}$.
\medskip

\subsection{Coaxial Line}
\label{sec:coax}

The cross-section is the ring-shaped geometric region bounded by two concentric circles, $a<r<b$ with conductors at $r=a$ and $r=b$. The separation of Eq.~\eqref{eq:circhelm} is the same as for the disk, $\psi=R(r)$ times $\cos(n\varphi)$ or $\sin(n\varphi)$, and gives Bessel's equation in the variable $s=\kc r$,

\begin{equation}
s^{2}u''+s\,u'+\left(s^{2}-n^{2}\right)u=0,
\qquad
u=A\,J_{n}(s)+B\,Y_{n}(s),
\label{eq:coaxode}
\end{equation}
where now both solutions are kept, since the axis is outside the domain. With $\chi=\kc a$ the two wall conditions of each family fix the ratio $B/A$ and the admissible values of $\chi$.

\subsubsection{TM Modes}

The generator $\Ez=u(\kc r)\cos(n\varphi)$ must vanish at both conductors. The inner condition $u(\chi)=0$ fixes the combination

\begin{equation}
u_{nm}(s)=J_n(s)\,Y_n(\chi)-Y_n(s)\,J_n(\chi),
\label{eq:coaxu}
\end{equation}
\smallskip

\noindent and the outer condition $u_{nm}(b\chi/a)=0$ is the characteristic equation

\begin{equation}
F_n\!\left(\chi;\frac{b}{a}\right)=J_n(\chi)\,Y_n\!\left(\frac{b\chi}{a}\right)-Y_n(\chi)\,J_n\!\left(\frac{b\chi}{a}\right)=0,
\label{eq:coaxTM}
\end{equation}
\medskip

With $\Ez=E_0u_{nm}(\chi r/a)\cos(n\varphi)$, Eqs.~\eqref{eq:transM} and~\eqref{eq:transBM} give

\begin{equation}
\begin{aligned}
E_{r}&=\frac{ik}{\kc}\,E_0\,u_{nm}'\!\left(\frac{\chi r}{a}\right)\cos(n\varphi),\\[6pt]
E_{\varphi}&=-\frac{ikn}{\kc^{2}r}\,E_0\,u_{nm}\!\left(\frac{\chi r}{a}\right)\sin(n\varphi),\\[6pt]
B_{r}&=\frac{i\omega\mu_{0}\eps\,n}{\kc^{2}r}\,E_0\,u_{nm}\!\left(\frac{\chi r}{a}\right)\sin(n\varphi),\\[6pt]
B_{\varphi}&=\frac{i\omega\mu_{0}\eps}{\kc}\,E_0\,u_{nm}'\!\left(\frac{\chi r}{a}\right)\cos(n\varphi),
\end{aligned}
\label{eq:coaxtmfields}
\end{equation}
\smallskip

\noindent where the prime denotes differentiation with respect to $s$, with $\kc=\chi/a$ and the dispersion relation $k^{2}=\omega^{2}\mu_{0}\eps-\chi_{nm}^{2}/a^{2}$. The sine branch follows by exchanging $\cos(n\varphi)$ and $\sin(n\varphi)$ with the sign of $E_{\varphi}$ and $B_{r}$ reversed, and only the cosine branch exists for $n=0$.

\subsubsection{TE Modes}

The generator $\Bz=v(\kc r)\cos(n\varphi)$ must have a vanishing radial derivative at both conductors. The inner condition $v'(\chi)=0$ fixes

\begin{equation}
v_{nm}(s)=J_n(s)\,Y_n'(\chi)-Y_n(s)\,J_n'(\chi),
\label{eq:coaxv}
\end{equation}
\smallskip

\noindent and the outer condition $v_{nm}'(b\chi/a)=0$ is the TE characteristic equation

\begin{equation}
G_n\!\left(\chi;\frac{b}{a}\right)=J_n'(\chi)\,Y_n'\!\left(\frac{b\chi}{a}\right)-Y_n'(\chi)\,J_n'\!\left(\frac{b\chi}{a}\right)=0,
\label{eq:coaxTE}
\end{equation}
\smallskip

\noindent which is Eq.~\eqref{eq:coaxTM} with every Bessel function replaced by its derivative. With $\Bz=B_0v_{nm}(\chi r/a)\cos(n\varphi)$ and $\Ez=0$, the same relations give

\begin{equation}
\begin{aligned}
B_{r}&=\frac{ik}{\kc}\,B_0\,v_{nm}'\!\left(\frac{\chi r}{a}\right)\cos(n\varphi),
&\qquad
B_{\varphi}&=-\frac{ikn}{\kc^{2}r}\,B_0\,v_{nm}\!\left(\frac{\chi r}{a}\right)\sin(n\varphi),
\\[2pt]
E_{r}&=-\frac{i\omega n}{\kc^{2}r}\,B_0\,v_{nm}\!\left(\frac{\chi r}{a}\right)\sin(n\varphi),
&\qquad
E_{\varphi}&=-\frac{i\omega}{\kc}\,B_0\,v_{nm}'\!\left(\frac{\chi r}{a}\right)\cos(n\varphi).
\end{aligned}
\label{eq:coaxtefields}
\end{equation}

\subsubsection{TEM Mode}
With $\kc=0$ the potential obeys $\lapt\Phi=0$ in the annulus with $\Phi=V_0$ at $r=a$ and $\Phi=0$ at $r=b$. The axisymmetric solution of Laplace's equation in the plane is $\Phi=A\ln r+B$, and the two conductor values give

\begin{equation}
\Phi=V_0\frac{\ln(b/r)}{\ln(b/a)},
\qquad
E_r=-\frac{\partial\Phi}{\partial r}=\frac{V_0}{r\ln(b/a)},
\qquad
B_\varphi=\frac{E_r}{c},
\qquad
\Ez=\Bz=0,
\label{eq:coaxTEM}
\end{equation}
with $k=\omega/c$, the voltage $V=\int_a^b E_r\,d r=V_0$, and the line current from the wall field,
\begin{equation}
I=\frac{1}{\mu_{0}}\oint B_\varphi\,r\,d\varphi=\frac{2\pi V_0}{\ln(b/a)}\sqrt{\frac{\eps}{\mu_{0}}}.
\label{eq:coaxZ0}
\end{equation}

This is the one TEM mode of the line, $q-1=1$ for its two conductors.

\subsubsection{SL Modes}

The generator $C$ carries Dirichlet data at both conductors, the same data as the TM generator, so the radial function is again $u_{nm}$ of Eq.~\eqref{eq:coaxu} and the roots are those of Eq.~\eqref{eq:coaxTM}. A scalar mode with amplitude $C_0$ is

\begin{equation}
C_{nm}(r,\varphi)=
C_0\left[
J_n\left(\chi\frac{r}{a}\right)Y_n(\chi)
-
Y_n\left(\chi\frac{r}{a}\right)J_n(\chi)
\right]
\begin{cases}
\cos(n\varphi),\\
\sin(n\varphi),
\end{cases}
\label{eq:coaxprofile}
\end{equation}
\smallskip

\noindent that is $C_0u_{nm}(\chi r/a)$ times the angular factor, and Eq.~\eqref{eq:transSL} gives
\begin{align}
E_r&=-\frac{i}{\omega\mu_{0}\eps}\,\frac{\chi}{a}\,C_0\,u_{nm}'\!\left(\frac{\chi r}{a}\right)
\begin{cases}\cos(n\varphi)\\ \sin(n\varphi)\end{cases},
\label{eq:coaxErho}\\
E_\varphi&=-\frac{i}{\omega\mu_{0}\eps}\,\frac{n}{r}\,C_0\,u_{nm}\!\left(\frac{\chi r}{a}\right)
\begin{cases}-\sin(n\varphi)\\ \phantom{-}\cos(n\varphi)\end{cases},
\label{eq:coaxEphi}\\
E_z&=\frac{k}{\omega\mu_{0}\eps}\,C_0\,u_{nm}\!\left(\frac{\chi r}{a}\right)
\begin{cases}\cos(n\varphi)\\ \sin(n\varphi)\end{cases},
\label{eq:coaxEz}
\end{align}

where $\chi/a$ in Eq.~\eqref{eq:coaxErho} is the chain-rule factor $ds/d r$. Since $u_{nm}(\chi)=u_{nm}(b\chi/a)=0$, $E_\varphi$ and $E_z$ vanish at both conductors while $E_r$ may not. No SL mode occurs at $\kc=0$, because the TEM field of Eq.~\eqref{eq:coaxTEM} has $C=0$ and the annular Dirichlet problem has no zero eigenvalue. Table~\ref{tab:cutoffs} lists the lowest roots for $b/a=3$. For $n\ge1$ the sine and cosine angular functions give a twofold degeneracy. As $b/a\to1$ the radial profile approaches a sinusoid across the gap, and the large-argument asymptotics of the Bessel cross-product zeros give $\chi_{nm}\sim m\pi a/(b-a)$~\cite{watson}, the ladder of a thin annulus.
\\

\FloatBarrier

Corollary~\ref{cor:complete} gives the general scalar field in the line as

\begin{equation}
\begin{gathered}
\begin{aligned}
C&=e^{-i\omega t}\sum_{n\ge0}\sum_{m\ge1}u_{nm}\!\left(\frac{\chi_{nm}r}{a}\right)\Big[\left(a^{+}_{nm}e^{ik_{nm}z}+a^{-}_{nm}e^{-ik_{nm}z}\right)\cos(n\varphi)\\
&\qquad\qquad\qquad+\left(b^{+}_{nm}e^{ik_{nm}z}+b^{-}_{nm}e^{-ik_{nm}z}\right)\sin(n\varphi)\Big],
\end{aligned}\\[4pt]
k_{nm}^{2}=\omega^{2}\mu_{0}\eps-\frac{\chi_{nm}^{2}}{a^{2}},
\end{gathered}
\label{eq:coaxsum}
\end{equation}

where $\chi_{nm}$ runs over the roots of Eq.~\eqref{eq:coaxTM} and $b^{\pm}_{0m}=0$. The forward and backward waves of one radial root carry independent cosine and sine amplitudes, and at cutoff the axial factors become $A_{nm}+B_{nm}z$ for the cosine partner and $D_{nm}+F_{nm}z$ for the sine partner. On an infinite guide boundedness sets $B_{nm}$ and $F_{nm}$ to zero. The classical TM and TE families have matching expansions over $u_{nm}$ and $v_{nm}$.

\subsection{L-Shaped Guide}
\label{sec:lshape}

The three guides above are separable, so a fair test of Theorem~\ref{thm:slspectrum} needs a cross-section that is not. The L-shaped guide is the unit square with its upper right quadrant removed,
\begin{equation}
\mathcal A_L=\operatorname{int}\,\overline{R_1\cup R_2},
\qquad
R_1=(0,1)\times(0,\tfrac12),
\qquad
R_2=(0,\tfrac12)\times(\tfrac12,1),
\label{eq:lshape}
\end{equation}
with six conducting wall segments and a re-entrant corner at $(\tfrac12,\tfrac12)$. The two rectangles join through the opening $\Gamma$ at $y=\tfrac12$, $0<x<\tfrac12$, which lies inside the domain defined by Eq.~\eqref{eq:lshape}. The remaining segment at $y=\tfrac12$, $\tfrac12<x<1$, is a conducting shelf.
\\

Some modes have simple sine or cosine product forms, but these do not form a complete basis or include the lowest modes. The full spectrum follows from matching the solutions in the two rectangles across $\Gamma$.
\\

The outer side has unit length, and the cutoff wavenumbers for other sizes scale inversely with that length. This is the classical L-shaped geometry at half scale, whose eigenvalues are known to many digits~\cite{foxhenricimoler,betcketrefethen}.
\\

Equation~\eqref{eq:genhelm} holds in each rectangle, and in each it separates as in Section~\ref{sec:rect}. In $R_1$ the walls $x=0$ and $x=1$ quantize the $x$ dependence to $\sin(n\pi x)$, and the bottom wall selects the $y$ dependence that vanishes at $y=0$. In $R_2$ the walls $x=0$ and $x=\tfrac12$ quantize it to $\sin(2m\pi x)$, and the top wall selects the dependence that vanishes at $y=1$. A general solution with Dirichlet data on the outer walls is therefore the pair of series
\begin{equation}
\psi_1=\sum_{n\ge1}a_n\sin(n\pi x)\,\frac{\sin(p_n y)}{p_n},
\qquad
\psi_2=\sum_{m\ge1}b_m\sin(2m\pi x)\,\frac{\sin\!\big(q_m(1-y)\big)}{q_m},
\label{eq:lseries}
\end{equation}
\begin{equation}
p_n^{2}=\kc^{2}-(n\pi)^{2},
\qquad
q_m^{2}=\kc^{2}-(2m\pi)^{2},
\label{eq:lpq}
\end{equation}
\smallskip

\noindent where a term with $p_n=i\eta$ is read as $\sinh(\eta y)/\eta$ and a term with $p_n=0$ as $y$, both the limits of the written form. The two series are not yet one mode. They become one when they agree across the opening and when $\psi_1$ satisfies the wall condition on the shelf, and it is those conditions that fix $\kc$.

\subsubsection{TM Modes}
With $\psi=\Ez$, the condition $\Ez=0$ holds on all six wall segments, which Eq.~\eqref{eq:lseries} already satisfies on the outer four. Across the opening $\Ez$ and $\partial\Ez/\partial y$ are continuous, since there is no wall there, and on the shelf $\Ez=0$,
\begin{equation}
\begin{aligned}
\psi_1(x,\tfrac12)&=\psi_2(x,\tfrac12),
\qquad
\frac{\partial\psi_1}{\partial y}(x,\tfrac12)=\frac{\partial\psi_2}{\partial y}(x,\tfrac12)
\qquad(0<x<\tfrac12),\\[14pt]
\psi_1(x,\tfrac12)&=0
\qquad(\tfrac12<x<1).
\end{aligned}
\label{eq:lmatch}
\end{equation}
Multiplying the first and third conditions together by $\sin(n\pi x)$ and integrating over $0<x<1$, and the second by $\sin(2m\pi x)$ and integrating over the opening, turns them into two sets of linear equations for the coefficients,
\begin{equation}
\begin{gathered}
\frac{a_n}{2}\,\frac{\sin(p_n/2)}{p_n}=\sum_{m\ge1}I_{nm}\,b_m\,\frac{\sin(q_m/2)}{q_m},\\[4pt]
\sum_{n\ge1}I_{nm}\,a_n\cos(p_n/2)+\frac{b_m}{4}\cos(q_m/2)=0,
\end{gathered}
\label{eq:lsystem}
\end{equation}
with the overlaps of the two sets of sines over the opening
\begin{equation}
\begin{aligned}
I_{nm}&=\int_{0}^{1/2}\sin(n\pi x)\sin(2m\pi x)\,dx\\
&=\frac12\left[\frac{\sin\big((n-2m)\pi/2\big)}{(n-2m)\pi}-\frac{\sin\big((n+2m)\pi/2\big)}{(n+2m)\pi}\right],
\end{aligned}
\label{eq:loverlap}
\end{equation}
the first term read as $\tfrac14$ when $n=2m$. Equation~\eqref{eq:lsystem} has a nonzero solution only at particular values of $\kc$, which are the TM cutoff wavenumbers, and its truncation to $N$ terms in each series gives them to any wanted accuracy as $N$ grows. The fields then follow from Eqs.~\eqref{eq:transM} and~\eqref{eq:transBM} with $\Ez=E_0\psi$, exactly as for the rectangle, $\Et=(ik/\kc^{2})\gradt\Ez$ and $\Bt=(i\omega\mu_{0}\eps/\kc^{2})\,\zhat\times\gradt\Ez$, each series differentiated term by term in its own rectangle.

\subsubsection{TE Modes}
With $\psi=\Bz$, the condition is $\partial\Bz/\partial n=0$ on all six segments, so the sines in Eq.~\eqref{eq:lseries} become cosines with $n,m\ge0$ and the $y$ factors become $\cos(p_n y)$ and $\cos(q_m(1-y))$. The same three conditions hold at $y=\tfrac12$, continuity of $\Bz$ and of $\partial\Bz/\partial y$ across the opening and $\partial\Bz/\partial y=0$ on the shelf, and projecting them with cosines gives the TE version of Eq.~\eqref{eq:lsystem}. One root must be discarded. The constant $\Bz$ satisfies every Neumann condition with $\kc=0$, but integrating Faraday's law over the cross-section gives $i\omega\int_{\cA}\Bz\,dA=\oint\vE\cdot d\boldsymbol\ell=0$ on a perfect conductor, so a constant $\Bz$ vanishes at every nonzero frequency and only the positive roots are TE modes. The fields follow from the same relations with $\Ez=0$, $\Bt=(ik/\kc^{2})\gradt\Bz$ and $\Et=-(i\omega/\kc^{2})\,\zhat\times\gradt\Bz$.

\subsubsection{TEM Mode}
The L-shape is simply connected, so the argument of Section~\ref{sec:rect} applies unchanged. A transverse potential harmonic in $\mathcal A_L$ and constant on its single boundary is constant throughout, and the guide carries no TEM mode.

\subsubsection{SL Modes}
With $\psi=C$, the wall condition $C=0$ is the TM condition, the interface conditions are continuity of $C$ and of $\partial C/\partial y$, and the shelf condition is $C=0$, so Eq.~\eqref{eq:lsystem} is the SL characteristic system as well, entry for entry. The SL cutoffs of the L-shaped guide are therefore its TM cutoffs, which is Theorem~\ref{thm:slspectrum} on a cross-section where the full spectrum is not available in closed form, and the SL fields are $\vE=-(i/(\omega\mu_{0}\eps))\grad C$ with $\vB=\mathbf0$ and $C=C_0\psi$, the same profile that gives $\Ez$ of the TM mode.

\subsubsection{Cutoffs}
The lowest cutoffs were computed in two independent ways, from Eq.~\eqref{eq:lsystem} truncated at sixty terms in each series, and from the five-point finite-difference form of Eq.~\eqref{eq:dirichleteigs},
\begin{equation}
\frac{4u_{ij}-u_{i+1,j}-u_{i-1,j}-u_{i,j+1}-u_{i,j-1}}{h^{2}}=\lambda\,u_{ij},
\qquad
h_{\mathrm D}=\frac1N,
\qquad
h_{\mathrm N}=\frac{1}{2N},
\label{eq:fivepoint}
\end{equation}
with $u=0$ on the wall for TM and SL and the missing neighbors reflected for TE. The TM and SL problem is posed on the vertex grid with spacing $h=h_{\mathrm D}$, and the TE problem on the cell-centered grid with spacing $h=h_{\mathrm N}$, so that both grids resolve the walls at $x=1/2$ and $y=1/2$ and the index $N$ labels the pair. The two methods give close agreement, although sixty terms do not resolve the last displayed digit of the lowest TM cutoff. The third Dirichlet mode is the exact product $\sin(2\pi x)\sin(2\pi y)$ with $\kc^{2}=8\pi^{2}$, and the third TE mode is the degenerate pair $\cos(2\pi x)$ and $\cos(2\pi y)$ with $\kc^{2}=4\pi^{2}$, the lowest of the elementary even-index members of either family. The L-shaped entries of Table~\ref{tab:cutoffs} are the converged values. For the lowest TM cutoff, the sixty-term matching calculation gives $6.20865$, the finite-difference values extrapolate to $6.2095$, and the reference value is approximately $6.2096$, the eigenvalue of Ref.~\cite{betcketrefethen} scaled to the unit outer side, which rounds to the value $6.210$ in Table~\ref{tab:cutoffs}. The lowest TE cutoff agrees with the reference value $2.4295$ of Kleefeld~\cite{kleefeld2021} for the same domain. The lowest modes converge more slowly than the elementary ones because their profiles behave as $r^{2/3}$ near the $270^{\circ}$ corner, the usual behaviour of an eigenfunction at a reentrant corner~\cite{grisvard}, with $r$ the distance from the corner. Figure~\ref{fig:lshape} shows the first mode of each family, the TE$_1$ profile of $\Bz$, which changes sign across the guide, and the TM$_1$ profile of $\Ez$, which is also the SL$_1$ profile of $C$.

\section{Cutoffs, Power and Wall Losses}
\label{sec:cutoffs}
\subsection{Cutoff Frequencies, Classical and Extended}

Table~\ref{tab:cutoffs} lists the lowest cutoffs of the four guides for every mode family in the dimensionless form each section uses. The TM and SL columns are identical entry for entry, which is Theorem~\ref{thm:slspectrum} in numbers, and the TE column shows where the classical spectrum differs. In the rectangular guide only the TE family carries entries with a zero index. In the circular guide and the coaxial line the identity $J_{0}'=-J_{1}$ makes every TE$_{0m}$ coincide with TM$_{1m}$ and SL$_{1m}$, and only the coaxial line carries a TEM mode, at zero.

{\setlength{\intextsep}{2pt plus 2pt minus 2pt}%
\begin{table}[!htbp]
\caption{Lowest cutoffs of the four canonical guides by family. Rectangular values are $\kc b$ for $a=2b$, circular values $\kc a$, coaxial values $\chi=\kc a$ for $b/a=3$ with the TE roots from Eq.~\eqref{eq:coaxTE} and the TM and SL roots from Eq.~\eqref{eq:coaxTM}, and L-shaped values $\kc$ for a unit outer side, listed as distinct values in increasing order so that the double TE root $2\pi$ appears once. Only the coaxial line carries a TEM mode, at zero. A blank means no mode with those indices.}
\label{tab:cutoffs}
\centering
{\setlength{\tabcolsep}{4.5pt}\renewcommand{\arraystretch}{0.9}
\begin{tabular}{@{}lccc@{\hspace{1.6em}}lccc@{}}
\toprule
\multicolumn{4}{@{}l}{\textit{Rectangular Guide, $a=2b$, $\kc b$}} & \multicolumn{4}{l@{}}{\textit{Circular Guide, $\kc a$}}\\
\cmidrule(r{1.6em}){1-4}\cmidrule(l){5-8}
$(n,m)$ & TE & TM & SL & $(n,m)$ & TE & TM & SL\\
\midrule
$(1,0)$ & 1.5708 & & & $(1,1)$ & 1.8412 & 3.8317 & 3.8317\\
$(0,1)$ & 3.1416 & & & $(0,1)$ & 3.8317 & 2.4048 & 2.4048\\
$(2,0)$ & 3.1416 & & & $(2,1)$ & 3.0542 & 5.1356 & 5.1356\\
$(1,1)$ & 3.5124 & 3.5124 & 3.5124 & $(3,1)$ & 4.2012 & 6.3802 & 6.3802\\
$(2,1)$ & 4.4429 & 4.4429 & 4.4429 & $(4,1)$ & 5.3176 & 7.5883 & 7.5883\\
$(3,0)$ & 4.7124 & & & $(1,2)$ & 5.3314 & 7.0156 & 7.0156\\
$(3,1)$ & 5.6636 & 5.6636 & 5.6636 & $(0,2)$ & 7.0156 & 5.5201 & 5.5201\\
$(0,2)$ & 6.2832 & & & & & & \\
$(4,0)$ & 6.2832 & & & & & & \\
\midrule
\multicolumn{4}{@{}l}{\textit{Coaxial Line, $b/a=3$, $\chi=\kc a$}} & \multicolumn{4}{l@{}}{\textit{L-Shaped Guide, Unit Outer Side, $\kc$}}\\
\cmidrule(r{1.6em}){1-4}\cmidrule(l){5-8}
$(n,m)$ & TE & TM & SL & \multicolumn{1}{c}{$j$} & TE & TM & SL\\
\midrule
$(1,1)$ & 0.5136 & 1.6356 & 1.6356 & \multicolumn{1}{c}{1} & 2.430 & 6.210 & 6.210\\
$(2,1)$ & 0.9775 & 1.8680 & 1.8680 & \multicolumn{1}{c}{2} & 3.760 & 7.797 & 7.797\\
$(3,1)$ & 1.3880 & 2.1887 & 2.1887 & \multicolumn{1}{c}{3} & 6.283 & 8.886 & 8.886\\
$(0,1)$ & 1.6356 & 1.5485 & 1.5485 & \multicolumn{1}{c}{4} & 6.750 & 10.87 & 10.87\\
$(4,1)$ & 1.7692 & 2.5520 & 2.5520 & \multicolumn{1}{c}{5} & 7.092 & 11.30 & 11.30\\
\botrule
\end{tabular}}
\end{table}}

\FloatBarrier

\subsection{Power and Energy}

\subsubsection{Modal Power}
\label{sec:power}
\begin{proposition}[Power, Energy per Unit Length, and Group Velocity]
\label{prop:powerenergy}
Assume $k_j\in\mathbb{R}$ and $k_j\neq0$, and normalize $\phi_j$ as in Eq.~\eqref{eq:dirichleteigs}. The cutoff value follows by continuity with $P_j=0$. The time-averaged axial flux of $\vS$ and density of $u$ per unit length, the power and energy of the mode in the form of Eq.~\eqref{eq:density}, are

\begin{equation}
P_j=\frac{k_j}{2\mu_{0}^{2}\eps\omega}|A_j|^2,
\qquad
U_j=\frac{1}{2\mu_{0}}|A_j|^2.
\label{eq:PU}
\end{equation}

Consequently,

\begin{equation}
\frac{P_j}{U_j}=\frac{k_j}{\omega\mu_{0}\eps}=\frac{d\omega}{dk_j}.
\label{eq:PoverU}
\end{equation}
\end{proposition}

\begin{proof}
For an SL mode, applying the time average of a product of real parts, $\avg{fg}=\tfrac12\operatorname{Re}\{fg^{*}\}$, to the scalar flux $C\vE/\mu_{0}$ gives

\begin{equation}
\avg{\vS}=\frac{1}{2\mu_{0}}\operatorname{Re}\{C\vE^{*}\}.
\end{equation}

Using Eq.~\eqref{eq:slfields} gives

\begin{equation}
P_j=\int_{\cA}\avg{S_z}\,dA
=\frac{k_j}{2\mu_{0}^{2}\eps\omega}|A_j|^2\int_{\cA}|\phi_j|^2\,dA.
\end{equation}
\smallskip

In the SL sector, $\avg{u}=\eps|\vE|^2/4+|C|^2/(4\mu_{0})$. Since $|\grad C|^{2}=|\gradt C|^{2}+k_j^{2}|C|^{2}$, Eq.~\eqref{eq:gradorth} gives

\begin{equation}
\int_{\cA}|\grad C_j|^2\,dA
=\left(\lambda_j+k_j^2\right)|A_j|^2
=\omega^{2}\mu_{0}\eps|A_j|^2.
\end{equation}
\smallskip

Because $|\vE|^{2}=|\grad C|^{2}/(\omega\mu_{0}\eps)^{2}$, the integrated electric contribution is
\[
\frac{\eps}{4}\,\frac{\omega^{2}\mu_{0}\eps}{(\omega\mu_{0}\eps)^{2}}|A_j|^2
=\frac{1}{4\mu_{0}}|A_j|^2.
\]
This equals the scalar contribution $|A_j|^2/(4\mu_{0})$, so $U_j=|A_j|^2/(2\mu_{0})$. Differentiating $\omega=c\sqrt{k_j^2+\lambda_j}$ gives Eq.~\eqref{eq:PoverU}.
\end{proof}

$P_j$ and $U_j$ describe the quadratic form of Eq.~\eqref{eq:density}. They also equal the integrated modal power and energy from Eq.~\eqref{eq:Lpair} for the choice $\Phi=0$. The choice $\vA=\mathbf0$ gives their negatives, as discussed in Section~\ref{sec:extenergy}.
\\

Below cutoff $k_j=i\alpha_j$ with $\alpha_j=\sqrt{\lambda_j-\omega^{2}\mu_{0}\eps}$, the mode decays as $e^{-\alpha_j z}$, and it carries no time-averaged power because $P_j$ is proportional to $\operatorname{Re}k_j$, exactly as for an evanescent TE or TM mode.
\\

Equation~\eqref{eq:PoverU} has the same form as the energy-transport relation for guided Maxwell modes~\cite{collin}. Here $P_j$ and $U_j$ are the flux and density of the quadratic form of Eq.~\eqref{eq:density}, not the physical energy flux and density of Eq.~\eqref{eq:Lpair}. In Eq.~\eqref{eq:PU}, $P_j$ carries the sign of $k_j$. Below cutoff $k_j$ is imaginary, and a single evanescent mode carries no net time-averaged axial power, so Eqs.~\eqref{eq:PU} and~\eqref{eq:PoverU} apply only to propagating modes.
\\

\begin{proposition}[Modal Power Orthogonality]
\label{prop:orthogonality}
For co-frequency propagating SL modes $C=\sum_j a_j\phi_j e^{i(k_j z-\omega t)}$, with real $k_j$ and one propagation direction for each transverse eigenfunction, the time-averaged axial flux of $\vS$ is

\begin{equation}
P=\sum_j\frac{k_j}{2\mu_{0}^{2}\eps\omega}\,|a_j|^2,
\label{eq:poweradd}
\end{equation}

so the modal contributions to this quadratic flux add without cross terms.
\end{proposition}

\begin{proof}
Using $\avg{\vS}=\tfrac{1}{2\mu_{0}}\operatorname{Re}\{C\vE^{*}\}$ and $\Ez=\sum_k(k_k/(\omega\mu_{0}\eps))a_k\phi_k e^{i(k_k z-\omega t)}$, the axial power is

\begin{equation*}
P=\int_{\cA}\avg{S_z}\,dA
=\frac{1}{2\mu_{0}^{2}\eps\omega}\operatorname{Re}\sum_{j,k}k_k\,a_j a_k^{*}\,
e^{i(k_j-k_k)z}\int_{\cA}\phi_j\phi_k^{*}\,dA.
\end{equation*}
\smallskip

Orthonormality, $\int_{\cA}\phi_j\phi_k^{*}\,dA=\delta_{jk}$, reduces the double sum to Eq.~\eqref{eq:poweradd}.
\end{proof}

If both $+k_j$ and $-k_j$ components of one transverse profile are present, the interference term in $C E_z^{*}$ is $-(2ik_j/(\omega\mu_{0}\eps))\,|\phi_j|^{2}\,\mathrm{Im}\{a_{+}a_{-}^{*}e^{2ik_j z}\}$, and the factor $|\phi_j|^{2}$ integrates to one over the cross-section. It is purely imaginary and does not affect the time-averaged power. Equation~\eqref{eq:poweradd} therefore still holds when each component enters with its signed $k_j$. A good but imperfect conductor can be modeled through its skin-depth wall loss. This breaks the otherwise identical response of the TM and SL sectors.

\subsubsection{Wall Losses}
\label{sec:wallloss}

\begin{proposition}[Wall Loss and the SL Branches]
\label{prop:impedance}
Let the wall be a good conductor of conductivity $\sigma_c$ and skin depth $\delta=\sqrt{2/(\omega\mu_c\sigma_c)}$, with $\delta$ small compared with the dimensions of the guide, and keep $C=0$ on the wall. The wall then absorbs from the parallel magnetic field of each mode the time-averaged power per unit area~\cite{jackson}

\begin{equation}
p_{\mathrm{wall}}=\frac{(\sigma_c\delta)^{-1}}{2\mu_{0}^{2}}\left|\vB_{\parallel}\right|^{2}.
\label{eq:pwall}
\end{equation}
\smallskip

Every propagating TE, TM and TEM mode has $\vB_{\parallel}\ne\mathbf0$ somewhere on the wall and, to leading order in wall loss, where attenuation is a small correction to propagation, is attenuated along $z$ at the rate

\begin{equation}
\alpha=\frac{(\sigma_c\delta)^{-1}}{4\mu_{0}^{2}P}\oint_{\bnd}\left|\vB_{\parallel}\right|^{2}dl>0,
\label{eq:alphageneral}
\end{equation}
\smallskip

\noindent with $P>0$ the power the lossless mode transports along $+z$, the proposition being stated for forward modes with $k_j>0$. A backward mode is its mirror image, with the attenuation measured along its own direction of propagation. Every SL mode has $\vB_{\parallel}=\mathbf0$ and $\vE_{\parallel}=\mathbf0$ on the wall, so the method of Eq.~\eqref{eq:pwall} takes nothing from it, it drives no parallel wall current, and within this model it keeps the cutoff and dispersion of the lossless guide. What the conductor does with $C$ itself is outside the model. A wall that couples the two sectors, as the alternative condition of Section~\ref{sec:extwall} does, would give an SL field a Maxwell part with $\vB_{\parallel}\neq\mathbf0$, which this method could attenuate. Within the model, wall loss separates each TM mode from the SL mode that shared its cutoff.
\end{proposition}

\begin{proof}
For a Maxwell mode the power entering the wall per unit length is $\oint p_{\mathrm{wall}}\,dl$. The transported power decays as $P(z)=P(0)e^{-2\alpha z}$, so $-dP/dz=2\alpha P$, and equating the two gives Eq.~\eqref{eq:alphageneral} with the fields of the lossless mode~\cite{jackson}. The integrand is positive somewhere on the wall for every classical mode. For TM modes $\vB_{\parallel}$ is proportional to $\partial\Ez/\partial n$, which cannot vanish along the whole wall for a nontrivial mode. For TE modes $\Bz$ itself is nonzero on the wall. For the TEM mode $B_{\varphi}$ is nonzero on both conductors. Hence $\alpha>0$.
\\

For an SL mode $\vB=\mathbf0$ everywhere, so $\vB_{\parallel}=\mathbf0$ and $p_{\mathrm{wall}}=0$. Its parallel electric field on the wall is $-(i/(\omega\mu_{0}\eps))\,\hat{\mathbf t}\cdot\gradt C$ and $\Ez=(k/(\omega\mu_{0}\eps))C$, and both vanish because $C=0$ along the wall. The conductor's wall relation is therefore satisfied with zero on both sides, the mode is unchanged, and its wall flux $\avg{\vS}\cdot\nhat=\operatorname{Re}\{C\vE^{*}\}\cdot\nhat/(2\mu_{0})$ vanishes with $C$.
\end{proof}
\medskip

The model is the good-conductor approximation of Ref.~\cite{jackson}, valid where the skin depth is small compared with the wall's dimensions and curvature radii, and it enters only through the Maxwell field on the wall. What happens to $C$ inside a real conductor is not modeled. Within the model every nonzero-frequency field splits into a lossy Maxwell part and a lossless SL part, and the statement that the TM mode attenuates while the SL mode does not holds within the model and not beyond it. That scalar-longitudinal waves are unconstrained by the skin effect, because they have no magnetic field to drive eddy currents in a conductor, was asserted by Reed and Hively~\cite{reedhively2020}. The proposition above is the closed-guide version of that statement, with the wall model made explicit. As a scale, a copper guide of radius $11.5$ mm, whose TM$_{01}$ cutoff lies at $10$ GHz, loses $0.086$ dB\,m$^{-1}$ on that mode at $15$ GHz, while the SL$_{01}$ mode that shares the cutoff loses nothing through this mechanism.

\subsection{Evanescent Behavior and Velocities}

Below cutoff, $k_j=i\alpha_j$ with $\alpha_j=(\lambda_j-\omega^{2}\mu_{0}\eps)^{1/2}$, and the SL field decays as

\begin{equation}
C=A_j\phi_j\,e^{-\alpha_j z},
\qquad
\Et=-\frac{i}{\omega\mu_{0}\eps}A_j e^{-\alpha_j z}\gradt\phi_j,
\qquad
\Ez=\frac{i\alpha_j}{\omega\mu_{0}\eps}A_j\phi_j e^{-\alpha_j z}.
\label{eq:evanescent}
\end{equation}

\newpage

The factor of $i$ between $\Ez$ and $C$ makes $\operatorname{Re}\{C\Ez^{*}\}=0$ pointwise. Thus

\begin{equation}
\avg{S_z}=0
\qquad\text{for a single evanescent SL branch,}
\label{eq:evanescentpower}
\end{equation}
\smallskip

\noindent which is the same result as for a single evanescent TE or TM mode and explains why only propagating terms appear in the power balance of Section~\ref{sec:matching}.
\\

For propagating branches, all four families share $\omega^{2}\mu_{0}\eps=k^{2}+\kc^{2}$, so
\begin{equation}
v_p=\frac{\omega}{k},
\qquad
v_g=\frac{d\omega}{dk}=\frac{k}{\omega\mu_{0}\eps},
\qquad
v_pv_g=1/(\mu_{0}\eps),
\label{eq:velocities}
\end{equation}
as for TE and TM modes. A ratio intrinsic to the SL mode uses $C$. Equation~\eqref{eq:rectEz} gives
\begin{equation}
\frac{\Ez}{C}=\frac{k}{\omega\mu_{0}\eps}=v_g,
\label{eq:slratio}
\end{equation}
so the ratio of axial electric field to scalar field equals the group velocity. Since $C$ has the dimensions of $\vB$, Eq.~\eqref{eq:slratio} is dimensionally similar to the plane-wave ratio $E/B$, with $v_g$ in place of $c$.

\section{Discontinuities and Mode Matching}
\label{sec:matching}

The guides above are uniform in $z$. A step in cross-section is the simplest discontinuity, and it shows how the wall condition on $C$ carries across a change of guide. Let two perfectly conducting guides share an axis, with cross-section $\cA_1$ for $z<0$ and $\cA_2$ for $z>0$, and $\cA_1\subset\cA_2$, so that the ring $\cA_2\setminus\cA_1$ at $z=0$ is a conducting step face.
\\

An SL field has $\vB=\mathbf0$ and $\vE=-(i/(\omega\mu_{0}\eps))\grad C$, so the field is fixed by $C$ alone and the junction conditions are conditions on $C$. Across the aperture the scalar and the axial electric field are continuous, and on the step face the wall condition of Section~\ref{sec:slsector} applies,
\begin{equation}
\left.C\right|_{0^-}=\left.C\right|_{0^+},
\qquad
\left.\frac{\partial C}{\partial z}\right|_{0^-}=\left.\frac{\partial C}{\partial z}\right|_{0^+}
\quad\text{on }\cA_1,
\qquad\qquad
C=0
\quad\text{on }\cA_2\setminus\cA_1,
\label{eq:matchaperture}
\end{equation}
where the second condition is continuity of $\Ez=-(i/(\omega\mu_{0}\eps))\,\partial C/\partial z$. Continuity of $C$ gives continuity of $\Et=-(i/(\omega\mu_{0}\eps))\gradt C$ across the aperture, and $C=0$ on the step face gives $\vE_{\parallel}=\mathbf0$ there, so nothing further is needed. Because the incident field has no magnetic part and the conditions involve only $C$, no TE or TM mode is excited. An incident SL mode scatters only into SL modes.
\\

The scattered field follows in the usual way. On each side $C$ is a sum of the guide's own SL modes, incident and reflected for $z<0$ and transmitted for $z>0$, each with its own axial wavenumber from Eq.~\eqref{eq:genhelm}, real above cutoff and positive imaginary below it so that every evanescent term decays away from the step. Multiplying the first condition of Eq.~\eqref{eq:matchaperture} by a mode of the wide guide and integrating over $\cA_2$, and the second by a mode of the narrow guide and integrating over $\cA_1$, turns the two conditions into a linear system for the reflected and transmitted amplitudes, in which only the overlap integrals of the two sets of modes over the aperture appear. This is Dirichlet mode matching in its standard form~\cite{collin,wexler}, and Proposition~\ref{prop:powerenergy} supplies the check that the powers carried by the propagating reflected and transmitted modes add up to the incident power. For a step in width at fixed height the factor $\sin(m\pi y/b)$ is common to both sides and cancels, so the overlaps reduce to elementary integrals of two sines.
\\

The difference from a TM step is in what is matched. TM matching uses the parallel electric and magnetic fields, while SL matching uses $C$ and $\partial C/\partial z$ with $C=0$ on the step face, so the two families scatter differently even though they share the same cutoffs in each uniform section.

\FloatBarrier

\section{Excitation of the Scalar Sector}
\label{sec:green}

The complete transverse basis from Theorem~\ref{thm:slspectrum} also solves the driven scalar problem and shows which sources excite each SL mode.
\\

For a time-harmonic source, Eq.~\eqref{eq:Cwave} becomes
\begin{equation}
\left(\nabla^{2}+\omega^{2}\mu_{0}\eps\right)C=-\mu_{0}\Lambda,
\qquad
\Lambda\equiv\frac{\partial\rho_{\mathrm{tot}}}{\partial t}+\dive\vJ_{\mathrm{tot}},
\label{eq:Cdriven}
\end{equation}
where $C|_{\bnd}=0$ and $\Lambda$ is the charge-conservation residual. Expand $C=\sum_j c_j(z)\phi_j(x,y)$ and use
\begin{equation}
\int_{\cA}\phi_i^{*}\phi_j\,dA=\delta_{ij},
\qquad
\int_{\cA}\gradt\phi_i^{*}\cdot\gradt\phi_j\,dA=\lambda_j\delta_{ij},
\label{eq:orthoset}
\end{equation}
where the second relation follows from the first by Green's identity and $\phi_j|_{\bnd}=0$. Each coefficient then satisfies
\begin{equation}
c_j''+k_j^{2}c_j=-\mu_{0}\Lambda_j(z),
\qquad
\Lambda_j(z)=\int_{\cA}\phi_j^{*}(x,y)\,\Lambda(x,y,z)\,dA.
\label{eq:modeODE}
\end{equation}
Away from all cutoffs, $k_j\ne0$. The Green function of the guide is the solution of
\begin{equation}
\left(\nabla^{2}+\omega^{2}\mu_{0}\eps\right)G(\mathbf r,\mathbf r')=-\delta(\mathbf r-\mathbf r'),
\qquad
G=0\ \text{on the wall},
\label{eq:greendef}
\end{equation}
that is outgoing at $z\to\pm\infty$. Expanding it in the same modes, each coefficient obeys $g''+k_j^{2}g=-\delta(z-z')$, whose outgoing solution is $g=(i/2k_j)e^{ik_j|z-z'|}$, so
\begin{equation}
G(\mathbf r,\mathbf r')
=\sum_j\frac{i}{2k_j}\,\phi_j(x,y)\,\phi_j^{*}(x',y')\,e^{ik_j\left|z-z'\right|},
\label{eq:greenfn}
\end{equation}
with $k_j$ chosen as in Eq.~\eqref{eq:genhelm}, so that below-cutoff terms decay away from the source. The scalar field of any source is then the integral of this kernel against the charge-conservation residual,
\begin{equation}
C(\mathbf r)=\mu_{0}\int G(\mathbf r,\mathbf r')\,\Lambda(\mathbf r')\,dV',
\label{eq:greensol}
\end{equation}
\smallskip

\noindent and beyond the source the amplitude of mode $j$, outgoing for real $k_j$ and decaying for $k_j=i\alpha_j$, is
\begin{equation}
a_j=\frac{i\mu_{0}}{2k_j}\int\phi_j^{*}(x',y')\,\Lambda(\mathbf r')\,e^{-ik_j z'}\,dV'.
\label{eq:excitation}
\end{equation}
\smallskip

Equations~\eqref{eq:greenfn} and~\eqref{eq:excitation} exclude exact cutoff. At $\omega=c\sqrt{\lambda_j}$, $k_j=0$ and Eq.~\eqref{eq:modeODE} becomes
\begin{equation}
c_j''(z)=-\mu_{0}\Lambda_j(z),
\label{eq:threshold}
\end{equation}
\smallskip

\noindent while the cutoff equation has no bounded outgoing solution of the Helmholtz form above. Its one-dimensional Green function has the usual $1/k_j$ threshold singularity, so that channel needs a separate limiting treatment. Equation~\eqref{eq:excitation} makes the source rule in Section~\ref{sec:discussion} explicit. For the outgoing solution with no incoming homogeneous scalar field, only $\Lambda$ drives the modal amplitudes. A charge-conserving source has $\Lambda=0$, so $a_j=0$ for every mode regardless of the current distribution. The corresponding free-space excitation rule appears in Ref.~\cite{rentzber2026}. Nonconserved charge, or a numerical method with a nonzero discrete continuity residual, couples according to the overlap of that residual with $\phi_j$. In the divergence-cleaning interpretation of Section~\ref{sec:cleaning}, Eq.~\eqref{eq:excitation} is the modal amplitude driven by a discrete continuity residual, and Eq.~\eqref{eq:greenfn} carries it along the guide. Even when $\Lambda=0$, an initial Gauss-law error or an error in the boundary data can excite the scalar field. To exclude both contributions, the scalar initial data must be zero, and no scalar excitation can enter through the boundaries.

\FloatBarrier

\section{Illustrations}
\label{sec:illustrations}

\begin{figure}[!htbp]
\centering
\includegraphics[width=0.92\linewidth]{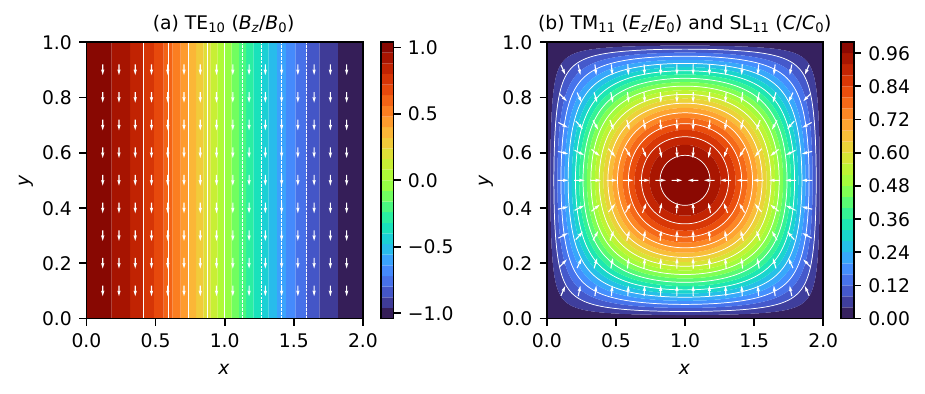}
\caption{Rectangular guide with $a=2b$. (a) TE$_{10}$. Filled contours show $\Bz/B_0$, and white arrows show $\zhat\times\gradt\Bz$, the direction of the transverse electric field. (b) TM$_{11}$ and SL$_{11}$, which share one profile. Filled contours show $\Ez/E_0$ and $C/C_0$, and white arrows show $\gradt C$, the direction of the transverse electric field of both modes.}
\label{fig:sl}
\end{figure}

Figure~\ref{fig:sl} shows field patterns in the rectangular guide with $a=2b$. Panel (a) shows TE$_{10}$. The filled contours represent $\Bz$, and the white arrows follow $\zhat\times\gradt\Bz$, showing the electric-field pattern given by Eq.~\eqref{eq:trans}.
\\

Panel (b) shows the shared profile of TM$_{11}$ and SL$_{11}$. By Theorem~\ref{thm:slspectrum}, the function $\sin(\pi x/a)\sin(\pi y/b)$ describes $\Ez$ for the TM mode and $C$ for the SL mode, up to their respective amplitudes. The arrows follow $\gradt C$ and point toward the maximum of the displayed profile. They show the spatial pattern of the electric-field component perpendicular to the guide axis for both modes.
\\

The field relations contain factors of $i$, so the arrows show spatial patterns rather than the field values at a particular instant. For the SL mode, the full electric field is proportional to $\grad C$. It also has a component parallel to the guide axis because $C$ varies with $z$. The TM mode has its own electric-field component $\Ez$ parallel to the axis, together with a magnetic field perpendicular to the electric-field arrows.

\FloatBarrier

\begin{figure}[!htbp]
\centering
\includegraphics[width=0.98\linewidth]{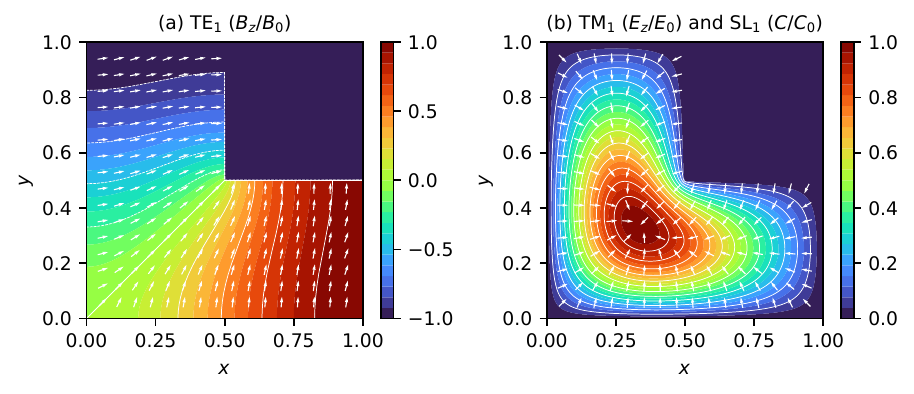}
\caption{First mode of each family in the L-shaped guide of Eq.~\eqref{eq:lshape} on the $N=400$ grids, with $h_{\mathrm N}=1/800$ in (a) and $h_{\mathrm D}=1/400$ in (b), each normalized to unit maximum. Panel (a) is the TE$_1$ profile of $\Bz$. Panel (b) is the TM$_1$ profile of $\Ez$, which by Theorem~\ref{thm:slspectrum} is also the SL$_1$ profile of $C$. White arrows show $\zhat\times\gradt\Bz$ in (a) and $\gradt\Ez$ in (b), the direction of the transverse electric field of each mode.}\label{fig:lshape}
\end{figure}

The two panels of Figure~\ref{fig:lshape} show the first mode of each family in the L-shaped guide of Section~\ref{sec:lshape}, computed on the $N=400$ grids of Eq.~\eqref{eq:fivepoint}, with $h_{\mathrm D}=1/400$ for the vertex grid of panel (b) and $h_{\mathrm N}=1/800$ for the cell-centered grid of panel (a). The TE$_1$ profile of $\Bz$ changes sign across the guide, negative in one arm and positive in the other with its zero line through the corner square, which is the odd symmetry of the lowest Neumann mode under the exchange of $x$ and $y$. The TM$_1$ profile of $\Ez$ is positive throughout, vanishes on every wall, and is also the SL$_1$ profile of $C$, since the two families share the Dirichlet problem. Near the re-entrant corner it behaves as $r^{2/3}$, as discussed there, so the profile itself vanishes at the corner while its gradient, and with it the transverse electric field of both modes, grows as $r^{-1/3}$.

\FloatBarrier

\newpage

\section{Scalar-Field Interpretation}
\label{sec:cleaning}

The additional terms introduced into the extended Gauss's and Amp\`ere's law can be physically interpreted as effective source densities,
\begin{equation}
\rho_{C}\equiv-\eps\,\pdt{C}, \qquad \vJ_{C}\equiv\frac{1}{\mu_{0}}\grad C,
\label{eq:effsources}
\end{equation}
because they enter Gauss's and Amp\`ere's laws in the same positions as charge and current densities. These definitions do not introduce additional charged matter. They express derivatives of the scalar field in source form. This distinction does not imply that the associated electric field cannot exert a force on matter.
\\

Using Eq.~\eqref{eq:Cwave}, the effective densities satisfy

\begin{equation}
\frac{\partial\rho_{C}}{\partial t}+\dive\vJ_{C}=-\eps\frac{\partial^{2}C}{\partial t^{2}}+\frac{1}{\mu_{0}}\nabla^{2}C=-\Lambda.
\end{equation}
Consequently,

\begin{equation}
\begin{aligned}
&\frac{\partial}{\partial t}\left(\rho_{\mathrm{tot}}+\rho_{C}\right)+\dive\left(\vJ_{\mathrm{tot}}+\vJ_{C}\right)\\
&\qquad=\Lambda-\frac{1}{\mu_{0}}\left(\mu_{0}\eps\frac{\partial^{2}C}{\partial t^{2}}-\nabla^{2}C\right)=0.
\end{aligned}
\label{eq:effcontinuity}
\end{equation}
\smallskip

This continuity relation follows from the extended field equations and applies to the combined effective sources. It does not require the matter charge and current to satisfy a separate continuity equation when $\Lambda\neq0$. The comparison with bound charge in a dielectric is limited to the way these terms enter the field equations. \\
\noindent Here, $\rho_{C}$ and $\vJ_{C}$ are defined from $C$ rather than from a material polarization.
\\

For a matter source that conserves charge, $\Lambda=0$, and Eq.~\eqref{eq:Cwave} becomes a homogeneous wave equation. Such a source does not generate $C$ through the bulk source term, but nonzero scalar waves can still be present through initial or boundary data. The conclusion $C=0$, and hence $\rho_{C}=\vJ_{C}=0$, additionally requires vanishing scalar initial data and no scalar excitation through the boundaries.
\\

The scalar-longitudinal family of Section~\ref{sec:slsector} has an irrotational electric field. Faraday's law then gives
\begin{equation}
\frac{\partial\vB}{\partial t}=-\curl\vE=0.
\end{equation}
\smallskip

\noindent Thus, for a mode with nonzero frequency, its magnetic field vanishes. The relation $\curl\vJ_{C}=\mu_{0}^{-1}\curl\grad C=0$ is consistent with this result, but does not by itself exclude an independently present transverse electromagnetic wave or a static magnetic background.
\\

In the empty interior of a guide, a scalar-longitudinal mode can therefore have nonzero $\rho_{C}$ and $\vJ_{C}$ even though $\rho_{\mathrm{tot}}=\vJ_{\mathrm{tot}}=0$ there. Their spatial profiles depend on the mode and may contain nodes. Any induced surface charge on the conductor, as discussed in Section~\ref{sec:slsector}, is a matter response at the boundary and is distinct from the effective density $\rho_{C}$ in the guide interior.
\\

The field-form wave equations retain their classical form. Repeating the curl computations leading to Eq.~\eqref{eq:fieldwave} with the extended system, the explicit terms in $C$ cancel, and Eq.~\eqref{eq:fieldwave} holds as written. Nevertheless, the allowed fields remain constrained by the extended first-order equations~\eqref{eq:eedfields}, which permit the additional scalar-longitudinal sector. In a vacuum region containing no matter sources, $\vE$, $\vB$, and $C$ each obey the vacuum wave equation.
\\

In regards to applications within computational physics, this field $C$ is also the scalar used in hyperbolic divergence cleaning. A Maxwell discretization that does not preserve Gauss's law can develop an error in $\dive\vE=\rho_{\mathrm{tot}}/\eps$, and hyperbolic divergence cleaning removes the error by adding a scalar $\psi$ that carries it away as a wave and damps it~\cite{munz2000,dedner2002}, since divergence error left in place produces spurious modes and wrong spectra~\cite{jiangwupovinelli}. With cleaning speed $v$ and damping time $\tau$ the system is
\begin{align}
\frac{\partial\vB}{\partial t}&=-\curl\vE,
\label{eq:cleanA}\\
\frac{\partial\vE}{\partial t}&=\frac{\curl\vB}{\mu_{0}\eps}-v^{2}\grad\psi-\frac{\vJ_{\mathrm{tot}}}{\eps},
\label{eq:cleanB}\\
\frac{\partial\psi}{\partial t}&=-\dive\vE+\frac{\rho_{\mathrm{tot}}}{\eps}-\frac{\psi}{\tau},
\label{eq:cleanC}
\end{align}
together with $\dive\vB=0$. Differentiating Eq.~\eqref{eq:cleanC} in time, substituting Eq.~\eqref{eq:cleanB}, and using $\dive\curl\vB=0$ gives
\begin{equation}
\nabla^{2}\psi-\frac{1}{v^{2}}\frac{\partial^{2}\psi}{\partial t^{2}}-\frac{1}{v^{2}\tau}\frac{\partial\psi}{\partial t}
=-\frac{1}{\eps v^{2}}\left(\frac{\partial\rho_{\mathrm{tot}}}{\partial t}+\dive\vJ_{\mathrm{tot}}\right),
\label{eq:cleantelegraph}
\end{equation}
which is the wave equation of $C$ with a different speed and a damping term, driven by the same charge-conservation residual. The scalar also contributes $\eps v^{2}\psi^{2}/2$ to the quadratic density of the cleaning system and $\eps v^{2}\psi\vE$ to its flux, which are the $C^{2}/(2\mu_{0})$ and $C\vE/\mu_{0}$ of Eq.~\eqref{eq:density} when $v=c$. With finite damping this quadratic norm is not conserved, and the damping term removes it at the rate $\eps v^{2}\psi^{2}/\tau$ per unit volume.
\\

With $\psi=0$ on a perfect conductor~\cite{munz2000,munzomnesschneider,crestetto2012}, a guided mode of angular frequency $\omega$ and axial wavenumber $k$ has the form $\psi=\phi_j(x,y)e^{i(kz-\omega t)}$, where $\phi_j$ is a Dirichlet eigenfunction of the cross-section with eigenvalue $\lambda_j$, $\lapt\phi_j+\lambda_j\phi_j=0$ and $\phi_j=0$ on the wall, exactly as in Eq.~\eqref{eq:dirichleteigs}. Substituting this form into the source-free Eq.~\eqref{eq:cleantelegraph} gives

\begin{equation}
k^{2}=\frac{\omega^{2}}{v^{2}}+\frac{i\omega}{v^{2}\tau}-\lambda_j.
\label{eq:cleandisp}
\end{equation}
\smallskip

The damping makes $k$ complex, so each mode decays along the guide. The transverse eigenvalues $\lambda_j$ of the cleaning scalar do not depend on $v$ or $\tau$, and the corresponding cutoff wavenumbers $\sqrt{\lambda_j}$ are listed in dimensionless form in Table~\ref{tab:cutoffs}. It has the TM cutoff frequencies themselves only at $v=c$ with no damping, where Eq.~\eqref{eq:cleandisp} becomes Eq.~\eqref{eq:cleanspectrum} below, and with damping the transition is not sharp.
\\

Now set

\begin{equation}
v=c,
\qquad
\tau^{-1}=0,
\qquad
\psi=C.
\label{eq:cleanident}
\end{equation}
\smallskip

Since $\eps c^{2}=1/\mu_{0}$, Eqs.~\eqref{eq:cleanA} to~\eqref{eq:cleanC} become Eq.~\eqref{eq:eedfields}, Eq.~\eqref{eq:cleantelegraph} becomes Eq.~\eqref{eq:Cwave}, and Eq.~\eqref{eq:cleandisp} becomes

\begin{equation}
k^{2}=\omega^{2}\mu_{0}\eps-\lambda_j,
\label{eq:cleanspectrum}
\end{equation}
\smallskip

\noindent which is Theorem~\ref{thm:slspectrum}. Every SL mode of Section~\ref{sec:geom} is therefore an exact mode of undamped speed-$c$ cleaning, with $\vB=\mathbf0$ and $\vE=-(i/(\omega\mu_{0}\eps))\grad\psi$. Nothing about the physical reality of $C$ enters.
\\

One exact solution serves as a benchmark for any $v$ and $\tau$. Close a length $L$ of guide with conducting end walls, so that $\psi=0$ there as well, and let $p$ be a positive integer counting the half-wavelengths along $z$. Then, with the same $\phi_j$ and $\lambda_j$, an amplitude $\psi_{0}$, and a ringing frequency $\Omega$,

\begin{equation}
\psi=\psi_{0}\,\phi_j(x,y)\sin\!\left(\frac{p\pi z}{L}\right)e^{-t/(2\tau)}\cos(\Omega t),
\qquad
\Omega^{2}=v^{2}\left[\lambda_j+\left(\frac{p\pi}{L}\right)^{2}\right]-\frac{1}{4\tau^{2}},
\label{eq:cleancavity}
\end{equation}
with $\vB=\mathbf0$ and, writing $\phi_{jp}=\phi_j(x,y)\sin(p\pi z/L)$ and $\lambda_{jp}=\lambda_j+(p\pi/L)^{2}$ for the bracket in $\Omega^{2}$,

\begin{equation}
\vE=\frac{\psi_{0}}{\lambda_{jp}}\,e^{-t/(2\tau)}\left[\frac{\cos(\Omega t)}{2\tau}-\Omega\sin(\Omega t)\right]\grad\phi_{jp},
\label{eq:cleanE}
\end{equation}
\smallskip

\noindent solves Eqs.~\eqref{eq:cleanA} to~\eqref{eq:cleanC} exactly when $\Omega^{2}>0$. The bracketed term in $\Omega^{2}$ is the total squared wavenumber of the standing mode, including contributions across the guide and along its axis. The final term accounts for damping.
\\

Each field equation can be checked directly. Faraday's law holds because $\vE$ is a gradient. The scalar equation follows from $\dive\vE=-\partial\psi/\partial t-\psi/\tau$, which is built into the solution. Amp\`ere's law requires $\partial\vE/\partial t=-v^{2}\grad\psi$. This follows when the amplitude satisfies the damped oscillator equation obtained by projecting Eq.~\eqref{eq:cleantelegraph} onto $\phi_{jp}$.
\\

A numerical simulation initialized with these fields should reproduce oscillations at frequency $\Omega$, with an amplitude that decays at the rate $1/(2\tau)$. When $\Omega^{2}<0$, the cosine and sine are replaced by hyperbolic functions with argument $(1/4\tau^{2}-v^{2}\lambda_{jp})^{1/2}\,t$. The mode then decays without oscillating. Its slower decay rate, which governs the long-time behavior when that contribution is present, is $1/(2\tau)-(1/4\tau^{2}-v^{2}\lambda_{jp})^{1/2}$.
\\

Setting $v=c$ and removing damping recovers the scalar cavity mode of Section~\ref{sec:slsector}. For a guide with the same wall conditions, undamped cleaning at $v=c$ cannot support a propagating scalar mode below the first TM cutoff. A numerical solution with a real axial wavenumber in that range would therefore indicate an error. Nonzero transmitted scalar flux alone does not indicate an error. Evanescent fields near sources or walls are allowed, and evanescent coupling through a short section can transmit flux. When $v<c$, the scalar cutoff $v\sqrt{\lambda_{1}}$ lies below the TM cutoff. With finite damping, there is no longer a sharp cutoff, as Eq.~\eqref{eq:cleandisp} shows.

\section{Discussion and Conclusions}
\label{sec:discussion}
\subsection{What the Results Say}

With $C=0$ on the wall, a perfectly conducting guide cannot support a freely propagating SL mode below its first TM cutoff, regardless of how it is excited. For propagating modes in the scalar sector alone, the conducting wall itself requires $C=0$. The scalar modes therefore follow the Dirichlet spectrum of the cross-section. Its eigenvalues are positive for every bounded cross-section, so the scalar family has no zero-cutoff mode and no counterpart to the TEM mode. Evanescent fields can still couple across a finite section of the guide below cutoff, just as they do in classical electrodynamics.
\\

Under this boundary condition, the classical TE, TM and TEM modes supported by the guide remain unchanged. The additional SL modes have no magnetic field, an electric field proportional to $\grad C$, and the same cutoffs as the TM modes. They therefore add a propagation channel only where a TM mode already propagates.
\\

Finite conductivity distinguishes the SL modes from their TM partners through wall losses. With $C=0$ and the skin-effect model of Section~\ref{sec:cutoffs}, the TM mode is attenuated, while the SL mode experiences no loss through that method. This result applies specifically to the wall condition and loss model used here.
\\

All of this holds for the condition $C=0$. At a conductor described by its constitutive laws, Section~\ref{sec:extwall}, the matching gives $\partial C/\partial n=0$ instead and the problem changes. Wherever the scalar varies along the wall, the scalar and Maxwell fields are coupled, and no field of the scalar sector alone propagates. Those coupled modes are the modes of such a guide, and finding their cutoffs requires a separate analysis, beyond the scope of this paper.
\\

The same equations describe the auxiliary scalar of undamped divergence cleaning at speed $c$, which is why the worked guides double as benchmarks, and the same wall condition carries into the closed cavity, where the scalar modes are the Dirichlet modes of the volume. What the paper does not do is model how the sector would be driven from outside, how a real conductor responds to $C$, what a dielectric interface does to it, how the field would be detected, or what physical energy and flux a field with $C\neq0$ carries, and each of those needs a model of its own. One thing can be said without one. By Eq.~\eqref{eq:Cwave} a charge-conserving source leaves $C$ obeying the homogeneous wave equation, and if $C$ and $\partial C/\partial t$ start at zero with $C=0$ on the walls and no scalar field comes in through the guide ends, or on an infinite guide no scalar radiation comes in, the uniqueness of solutions of the wave equation with these data keeps $C$ at zero for all time. Ordinary sources acting on ordinary initial fields therefore never populate the sector, which is why the guides of everyday use show no trace of it.

\subsection{Relation to Prior Work}

The idea of treating $C$ as a dynamical field appears in early work by Ohmura~\cite{ohmura} and in a formally related form in Stueckelberg theory~\cite{stueckelberg,stueckelberg2}. Aharonov and Bohm developed this extension in their 1963 paper on the role of the potentials in quantum theory~\cite{aharonovbohm}. They added to the electromagnetic Lagrangian a term quadratic in the four-divergence of the potential, which gives that combination its own wave equation. The combination is proportional to $C$, which is why the theory is also known as Aharonov-Bohm electrodynamics~\cite{modanese2017,minotti2021}. Later studies often introduce the scalar directly through Eq.~\eqref{eq:Cdef}~\cite{vanvlaenderen,hively2012,hivelyloebl,modanese2017,minotti2021,reedhively2020}. Its connection to gauge conditions, whose history is reviewed in Ref.~\cite{jacksonokun}, can also be expressed through a four-vector Helmholtz decomposition. Relaxing the Lorenz condition allows an additional four-gradient contribution to the potential~\cite{woodside1999,woodside2009}, and related cosmological models use this freedom to obtain a propagating scalar~\cite{jimenezmaroto}.
\\
 
The same mathematical structure appears in computational electromagnetics, where an auxiliary scalar is used to control errors in Gauss's law. Marder introduced a diffusive correction~\cite{marder1987}, while the hyperbolic Maxwell method of Ref.~\cite{munz2000} allows divergence errors to propagate as waves. Related approaches include generalized-Lagrange-multiplier methods for magnetohydrodynamics~\cite{dedner2002} and divergence control in finite-element Maxwell methods~\cite{monk2003}. For hyperbolic cleaning without damping and with propagation speed $c$, the auxiliary scalar obeys the same homogeneous wave equation as $C$. The scalar modes studied here can therefore also describe divergence-cleaning modes when the boundary conditions agree.
\\
 
That agreement at the boundary is essential. In a bounded region, the scalar wave equation must be supplied with a boundary condition, and the extended Maxwell equations allow more than one choice. Fornet and Mouysset examine both the initial and boundary conditions of time-dependent divergence cleaning, building on their earlier analysis~\cite{fornet2012}. Their rectangular-guide example uses a zero scalar wall value but admits non-Maxwell waves when the scalar initial data are nonzero. With compatible initial data and charge-conserving sources, they identify an extension that recovers the Maxwell solutions and prove its existence and uniqueness by a perturbation argument~\cite{fornetmouysset}. Their analysis shows why both the scalar wall condition and the initial data must be specified.
\\
 
Lastly, the material-interface problem of Minotti and Modanese~\cite{minotti2023} determines the wall condition at a physical conductor. Under locally conserved charge, with their constitutive relations and plane-wave potentials, their Section~2.4 finds that the scalar cannot propagate inside a dielectric or an ohmic conductor, and that the transmitted longitudinal part is a pure-potential wave carrying no field. Matching the solutions on the two sides of the surface, their Eq.~(21), gives $\partial C/\partial n=0$ on the vacuum side, and at oblique incidence the reflected scalar wave is accompanied by an ordinary electromagnetic wave. Section~\ref{sec:extwall} reproduces their Eq.~(21) in field form, obtains the same condition from surface charge conservation, and shows that under it no purely SL mode propagates in a uniform guide. The guided modes derived in this paper are those of the problem with $C=0$ on the wall, the condition under which the sectors separate and the one used in divergence cleaning, while the coupled modes of a guide bounded by their conductor are left for a separate analysis. A thin layer inside the conductor, proposed by one of the authors as the realization of the matching, and the constitutive relation that holds within it are noted in Section~\ref{sec:extwall} and left to a separate paper.

\FloatBarrier
\backmatter

\section*{Statements and Declarations}
\noindent\textbf{Funding}\quad No funding was received for this study.

\medskip
\noindent\textbf{Competing Interests}\quad The authors declare no competing interests.

\medskip
\noindent\textbf{Data Availability}\quad The data generated in this study are the tabulated values in the manuscript, which follow from the equations given.

\medskip
\noindent\textbf{Code Availability}\quad A Python script was used to compute modes and prepare figures. The definitions, derivations, inputs, and conclusions are fully stated in the manuscript and do not depend on access to the script. The Python script is however available upon request.

\medskip
\noindent\textbf{Author Contributions}\quad N.R. formulated the boundary-value problem, carried out the derivations and computations, and wrote the manuscript. F.M. and G.M. developed the formulation of extended electrodynamics in material media and its energy law on which this work is based, derived the scalar wall condition at a physical conductor, and revised the manuscript. J.H. contributed to the development of the manuscript. A.P. reviewed and analyzed the manuscript.

\newpage

\FloatBarrier

\end{document}